\pdfoutput=1
\documentclass[12pt,a4paper]{article}

\usepackage{ifthen} 
\newboolean{pdflatex}
\setboolean{pdflatex}{true} 

\newboolean{articletitles}
\setboolean{articletitles}{true} 

\newboolean{uprightparticles}
\setboolean{uprightparticles}{false} 

\def\paperauthors{LHCb collaboration} 
\def\paperasciititle{First observation of CP violation and measurement of polarization in B+ -> rho(770)0 Kstar(892)+ decays} 
\def\papertitle{First observation of \CP violation and measurement of polarization in $\Bp\to\rho(770)^0\Kstar(892)^+$ decays} 
\def\paperkeywords{{High Energy Physics}, {LHCb}} 
\def\papercopyright{\the\year\ CERN for the benefit of the LHCb collaboration} 
\def\paperlicence{CC BY 4.0 licence}
\def\paperlicenceurl{https://creativecommons.org/licenses/by/4.0/}

\newif\ifEnableSectionTOCLinks
\EnableSectionTOCLinksfalse 

\usepackage[top=1in, bottom=1.25in, left=1in, right=1in]{geometry}

\usepackage{microtype}
\usepackage{lineno}  
\usepackage{xspace} 
\usepackage{caption} 

\usepackage{graphicx}  
\usepackage{color}
\usepackage{colortbl}
\graphicspath{{./figs/}} 

\usepackage{amsmath} 
\usepackage{amssymb}
\usepackage{amsfonts}
\usepackage{upgreek} 

\newcommand*\patchAmsMathEnvironmentForLineno[1]{%
\expandafter\let\csname old#1\expandafter\endcsname\csname #1\endcsname
\expandafter\let\csname oldend#1\expandafter\endcsname\csname
end#1\endcsname
 \renewenvironment{#1}%
   {\linenomath\csname old#1\endcsname}%
   {\csname oldend#1\endcsname\endlinenomath}%
}
\newcommand*\patchBothAmsMathEnvironmentsForLineno[1]{%
  \patchAmsMathEnvironmentForLineno{#1}%
  \patchAmsMathEnvironmentForLineno{#1*}%
}
\AtBeginDocument{%
\patchBothAmsMathEnvironmentsForLineno{equation}%
\patchBothAmsMathEnvironmentsForLineno{align}%
\patchBothAmsMathEnvironmentsForLineno{flalign}%
\patchBothAmsMathEnvironmentsForLineno{alignat}%
\patchBothAmsMathEnvironmentsForLineno{gather}%
\patchBothAmsMathEnvironmentsForLineno{multline}%
\patchBothAmsMathEnvironmentsForLineno{eqnarray}%
}

\usepackage[pdftex,
            pdfauthor={\paperauthors},
            pdftitle={\paperasciititle},
            pdfkeywords={\paperkeywords}]{hyperref}
\usepackage{hyperxmp}
\hypersetup{
    pdfcopyright={Copyright (C) \papercopyright},
    pdflicenseurl={\paperlicenceurl}
}

\usepackage[colorinlistoftodos,textsize=scriptsize]{todonotes}

\usepackage[bottom,flushmargin,hang,multiple]{footmisc}

\usepackage[all]{hypcap} 

\usepackage{xspace}
\usepackage{upgreek}

\def\lhcb   {\mbox{LHCb}\xspace}

\def\babar  {\mbox{BaBar}\xspace}

\def\MagUp {\mbox{\em Mag\kern -0.05em Up}\xspace}

\ifthenelse{\boolean{uprightparticles}}%
{

 \def\Ppi         {\ensuremath{\uppi}\xspace}
 
 \def\Prho        {\ensuremath{\uprho}\xspace}

 \def\PDelta      {\ensuremath{\Delta}\xspace}
 \def\PXi         {\ensuremath{\Xi}\xspace}
 \def\PLambda     {\ensuremath{\Lambda}\xspace}
 \def\PSigma      {\ensuremath{\Sigma}\xspace}
 \def\POmega      {\ensuremath{\Omega}\xspace}
 \def\PUpsilon    {\ensuremath{\Upsilon}\xspace}
 \let\oldPi\Pi
 \def\PPi         {\ensuremath{\oldPi}\xspace}

 \def\PB      {\ensuremath{\mathrm{B}}\xspace}
 \def\PD      {\ensuremath{\mathrm{D}}\xspace}

 \def\PK      {\ensuremath{\mathrm{K}}\xspace}
 \def\Pp      {\ensuremath{\mathrm{p}}\xspace}

 \def\Ps      {\ensuremath{\mathrm{s}}\xspace}

 \def\thebaroffset{0.0em}
}
{

 \def\Ppi         {\ensuremath{\pi}\xspace}
 
 \def\Prho        {\ensuremath{\rho}\xspace}

 \mathchardef\PDelta="7101
 \mathchardef\PXi="7104
 \mathchardef\PLambda="7103
 \mathchardef\PSigma="7106
 \mathchardef\POmega="710A
 \mathchardef\PUpsilon="7107
 \mathchardef\PPi="7105
 \def\PB      {\ensuremath{B}\xspace}
 \def\PD      {\ensuremath{D}\xspace}

 \def\PK      {\ensuremath{K}\xspace}
 \def\Pp      {\ensuremath{p}\xspace}

 \def\Ps      {\ensuremath{s}\xspace}

 \def\thebaroffset{0.18em}
}
\newcommand{\offsetoverline}[2][\thebaroffset]{\kern #1\overline{\kern -#1 #2}}%

\makeatletter
\ifcase \@ptsize \relax
  \newcommand{\miniscule}{\@setfontsize\miniscule{4}{5}}
\or
  \newcommand{\miniscule}{\@setfontsize\miniscule{5}{6}}
\or
  \newcommand{\miniscule}{\@setfontsize\miniscule{5}{6}}
\fi
\makeatother

\DeclareRobustCommand{\optbar}[1]{\shortstack{{\miniscule (\rule[.5ex]{1.25em}{.18mm})}
  \\ [-.7ex] $#1$}}

\def\squark    {{\ensuremath{\Ps}}\xspace}

\def\pion   {{\ensuremath{\Ppi}}\xspace}

\def\pip    {{\ensuremath{\pion^+}}\xspace}
\def\pim    {{\ensuremath{\pion^-}}\xspace}
\def\pipm   {{\ensuremath{\pion^\pm}}\xspace}
\def\pimp   {{\ensuremath{\pion^\mp}}\xspace}

\def\rhomeson {{\ensuremath{\Prho}}\xspace}
\def\rhoz     {{\ensuremath{\rhomeson^0}}\xspace}
\def\rhop     {{\ensuremath{\rhomeson^+}}\xspace}
\def\rhom     {{\ensuremath{\rhomeson^-}}\xspace}

\def\kaon    {{\ensuremath{\PK}}\xspace}

\def\KorKbar {\kern \thebaroffset\optbar{\kern -\thebaroffset \PK}{}\xspace}

\def\KS      {{\ensuremath{\kaon^0_{\mathrm{S}}}}\xspace}

\def\Kstarz  {{\ensuremath{\kaon^{*0}}}\xspace}

\def\Kstar   {{\ensuremath{\kaon^*}}\xspace}

\def\Kstarp  {{\ensuremath{\kaon^{*+}}}\xspace}
\def\Kstarm  {{\ensuremath{\kaon^{*-}}}\xspace}

\def\Dbar    {{\ensuremath{\offsetoverline{\PD}}}\xspace}
\def\D       {{\ensuremath{\PD}}\xspace}

\def\DorDbar {\kern \thebaroffset\optbar{\kern -\thebaroffset \PD}\xspace}
\def\Dz      {{\ensuremath{\D^0}}\xspace}
\def\Dzb     {{\ensuremath{\Dbar{}^0}}\xspace}
\def\Dp      {{\ensuremath{\D^+}}\xspace}
\def\Dm      {{\ensuremath{\D^-}}\xspace}

\def\DpDm    {\ensuremath{\Dp {\kern -0.16em \Dm}}\xspace}

\def\B       {{\ensuremath{\PB}}\xspace}

\def\BorBbar {\kern \thebaroffset\optbar{\kern -\thebaroffset \PB}\xspace}
\def\Bz      {{\ensuremath{\B^0}}\xspace}

\def\Bd      {{\ensuremath{\B^0}}\xspace}

\def\BdorBdbar {\kern \thebaroffset\optbar{\kern -\thebaroffset \Bd}\xspace}
\def\Bu      {{\ensuremath{\B^+}}\xspace}
\def\Bub     {{\ensuremath{\B^-}}\xspace}
\def\Bp      {{\ensuremath{\Bu}}\xspace}
\def\Bm      {{\ensuremath{\Bub}}\xspace}
\def\Bpm     {{\ensuremath{\B^\pm}}\xspace}

\def\Bs      {{\ensuremath{\B^0_\squark}}\xspace}

\def\BsorBsbar {\kern \thebaroffset\optbar{\kern -\thebaroffset \Bs}\xspace}

\def\Y#1S{\ensuremath{\PUpsilon{(#1S)}}\xspace}

\def\proton      {{\ensuremath{\Pp}}\xspace}

\def\Lz          {{\ensuremath{\PLambda}}\xspace}

\def\LorLbar     {\kern \thebaroffset\optbar{\kern -\thebaroffset \PLambda}\xspace}

\newcommand{\decay}[2]{\mbox{\ensuremath{#1\!\to #2}}\xspace}

\def\to                 {\ensuremath{\rightarrow}\xspace}

\def\CP                {{\ensuremath{C\!P}}\xspace}

\def\AT#1     {\ensuremath{A_{\mathrm{T}}^{#1}}\xspace}           

\def\C#1      {\ensuremath{\mathcal{C}_{#1}}\xspace}                       
\def\Cp#1     {\ensuremath{\mathcal{C}_{#1}^{'}}\xspace}                    
\def\Ceff#1   {\ensuremath{\mathcal{C}_{#1}^{\mathrm{(eff)}}}\xspace}        
\def\Cpeff#1  {\ensuremath{\mathcal{C}_{#1}^{'\mathrm{(eff)}}}\xspace}       
\def\Ope#1    {\ensuremath{\mathcal{O}_{#1}}\xspace}                       
\def\Opep#1   {\ensuremath{\mathcal{O}_{#1}^{'}}\xspace}                    

\newcommand{\aunit}[1]{\ensuremath{\text{\,#1}}}

\newcommand{\tev}{\aunit{Te\kern -0.1em V}\xspace}
\newcommand{\gev}{\aunit{Ge\kern -0.1em V}\xspace}
\newcommand{\mev}{\aunit{Me\kern -0.1em V}\xspace}
\newcommand{\kev}{\aunit{ke\kern -0.1em V}\xspace}
\newcommand{\ev}{\aunit{e\kern -0.1em V}\xspace}

\newcommand{\mevc}{\ensuremath{\aunit{Me\kern -0.1em V\!/}c}\xspace}
\newcommand{\gevc}{\ensuremath{\aunit{Ge\kern -0.1em V\!/}c}\xspace}
\newcommand{\mevcc}{\ensuremath{\aunit{Me\kern -0.1em V\!/}c^2}\xspace}
\newcommand{\gevcc}{\ensuremath{\aunit{Ge\kern -0.1em V\!/}c^2}\xspace}

\def\fb   {\ensuremath{\aunit{fb}}\xspace}
\def\invfb   {\ensuremath{\fb^{-1}}\xspace}

\newcommand{\stat}{\aunit{(stat)}\xspace}
\newcommand{\syst}{\aunit{(syst)}\xspace}

\def\deriv {\ensuremath{\mathrm{d}}}

\def\gsim{{~\raise.15em\hbox{$>$}\kern-.85em
          \lower.35em\hbox{$\sim$}~}\xspace}
\def\lsim{{~\raise.15em\hbox{$<$}\kern-.85em
          \lower.35em\hbox{$\sim$}~}\xspace}

\newcommand{\Real}{\ensuremath{\mathcal{R}e}\xspace}
\newcommand{\Imag}{\ensuremath{\mathcal{I}m}\xspace}

\def\pt         {\ensuremath{p_{\mathrm{T}}}\xspace}

\def\tell1  {TELL1\xspace}
\def\ukl1   {UKL1\xspace}

\newcommand{\ie}{\mbox{\itshape i.e.}\xspace}

\newcommand{\vs}{\mbox{\itshape vs.}\xspace}

\newcommand{\lhcborcid}[1]{\href{https://orcid.org/#1}{\hspace*{0.1em}\raisebox{-0.45ex}{\includegraphics[width=1em]{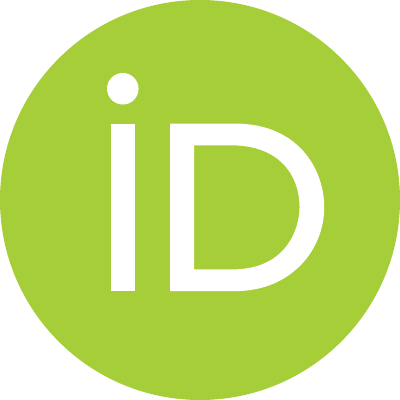}}}}

\hypersetup{
  colorlinks   = true, 
  urlcolor     = blue, 
  linkcolor    = blue, 
  citecolor    = red   
}

\ifEnableSectionTOCLinks
    \usepackage[explicit]{titlesec} 
    
    \let\oldcontentsline\contentsline
    \renewcommand

    \titleformat{\section}{\normalfont\Large\bf}{\hyperlink{tocsection.\thesection}{{\thesection} \parbox[t]{\dimexpr\textwidth-1pc}{#1}}}{1pc}{}

    \titleformat{\subsection}{\normalfont\bf}{\hyperlink{tocsubsection.\thesubsection}{{\thesubsection} \parbox[t]{\dimexpr\textwidth-1pc}{#1}}}{1pc}{}

    \titleformat{name=\section,numberless}[display]{}{}{0pt}{\normalfont\Huge\bfseries #1}
\fi

\usepackage{cite} 
\usepackage{mciteplus}

\usepackage{longtable} 
\def\ButoRhoKst {\ensuremath{\decay{\Bp}{\rho^0\Kstar^+}}\xspace}
\def\BtoVV {\ensuremath{\B\to VV}\xspace}

\def\myRho {\ensuremath{\rhoz}\xspace}
\def\myKst {\ensuremath{\Kstarp}\xspace}
\def\Azero {\ensuremath{A_0}\xspace}
\def\Aperp {\ensuremath{A_\perp}\xspace}
\def\Apara {\ensuremath{A_\parallel}\xspace}
\def\fLResult {{\ensuremath{0.720 \pm 0.028\stat \pm 0.009\syst}\xspace}} 
\def\ARhoKstResult {{\ensuremath{0.507 \pm 0.062\stat \pm 0.024\syst}\xspace}}

\def\fLResultSimple {{\ensuremath{0.720 \pm 0.028 \pm 0.009}\xspace}} 
\def\ARhoKstResultSimple {{\ensuremath{0.507 \pm 0.062 \pm 0.024}\xspace}} 
\def\ARhoKstLResultSimple {{\ensuremath{0.664 \pm 0.083 \pm 0.029}\xspace}} 
\def\AphaRhoKstLResultSimple {{\ensuremath{0.720 \pm 0.177 \pm 0.048}\xspace}}

\def\fLpResultSimple {{\ensuremath{0.491 \pm 0.083 \pm 0.025}\xspace}} 
\def\fLmResultSimple {{\ensuremath{0.794 \pm 0.025 \pm 0.006}\xspace}} 

\def\ARhoKstParaResultSimple {{\ensuremath{-0.063 \pm 0.137 \pm 0.027}\xspace}} 
\def\AphaRhoKstParaResultSimple {{\ensuremath{0.477 \pm 0.187 \pm 0.114}\xspace}} 
\def\ARhoKstPerpResultSimple {{\ensuremath{0.284 \pm 0.140 \pm 0.051}\xspace}} 
\def\AphaRhoKstPerpResultSimple {{\ensuremath{0.412 \pm 0.180 \pm 0.122}\xspace}} 

\def\AzeroVV {\ensuremath{A_0}\xspace}
\def\AperpVV {\ensuremath{A_\perp}\xspace}
\def\AparaVV {\ensuremath{A_\parallel}\xspace}
\def\AzeroVVbar {\ensuremath{\offsetoverline{A}_0}\xspace}
\def\AperpVVbar {\ensuremath{\offsetoverline{A}_\perp}\xspace}

\def\AiVV {\ensuremath{A_\lambda}\xspace}
\def\AiVVbar {\ensuremath{\offsetoverline{A}_\lambda}\xspace}

\def\AzeroVVbarbis {\ensuremath{\offsetoverline{A}{}^*_0}\xspace}

\def\AparaVVbarbis {\ensuremath{\offsetoverline{A}{}^*_\parallel}\xspace}
\def\AiVVbarbis {\ensuremath{\offsetoverline{A}{}^*_\lambda}\xspace}
\usepackage{booktabs}
\usepackage{multirow}
\usepackage{comment}
\begin{document}

\renewcommand{\thefootnote}{\fnsymbol{footnote}}
\setcounter{footnote}{1}
\begin{titlepage}
\pagenumbering{roman}

\vspace*{-1.5cm}
\centerline{\large EUROPEAN ORGANIZATION FOR NUCLEAR RESEARCH (CERN)}
\vspace*{1.5cm}
\noindent
\begin{tabular*}{\linewidth}{lc@{\extracolsep{\fill}}r@{\extracolsep{0pt}}}
\ifthenelse{\boolean{pdflatex}}
{\vspace*{-1.5cm}\mbox{\!\!\!\includegraphics[width=.14\textwidth]{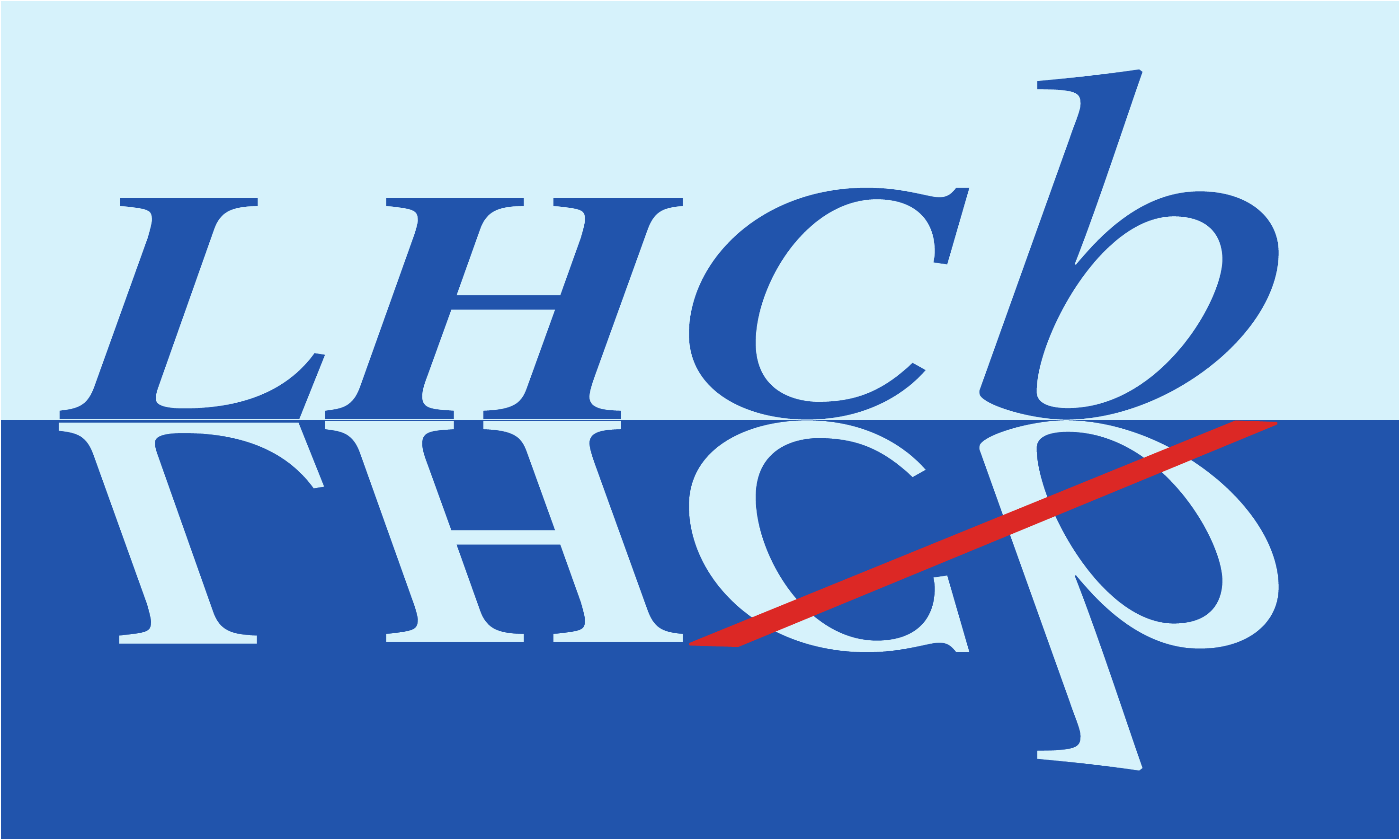}} & &}%
{\vspace*{-1.2cm}\mbox{\!\!\!\includegraphics[width=.12\textwidth]{figs/lhcb-logo.pdf}} & &}%
\\
 & & CERN-EP-2025-171                                                                 \\  
 & & LHCb-PAPER-2025-026 \\  
 & & April 10, 2026 \\ 
 & & \\
\end{tabular*}

\vspace*{4.0cm}

{\normalfont\bfseries\boldmath\huge
\begin{center}
  \papertitle 
\end{center}
}

\vspace*{2.0cm}

\begin{center}
\paperauthors\footnote{Authors are listed at the end of this paper.}
\end{center}

\vspace{\fill}

\begin{abstract}
  \noindent  An amplitude analysis of the \mbox{$\Bp\to(\pip\pim)(\KS\pip)$} decay  is performed in the mass regions \mbox{$0.30<m_{\pip\pim}<1.10\gevcc$} and \mbox{$0.75<m_{\KS\pip}<1.20\gevcc$}, using $pp$ collision data  recorded with the LHCb detector corresponding to an integrated luminosity of $9\invfb$. The polarization fractions and $\CP$ asymmetries for \mbox{$\Bp\to\rho(770)^0\Kstar(892)^+$} decays are measured.
  Violation of the $\CP$ symmetry in the decay \mbox{$\Bp\to\rho(770)^0\Kstar(892)^+$} is observed for the first time, with a significance exceeding nine standard deviations. The $\CP$ asymmetry is measured to be \mbox{$\mathcal{A}_{\CP} = \ARhoKstResult$} and the $\CP$-averaged longitudinal polarization fraction of \mbox{$f_L = \fLResult$}.
  The measurements help to shed light on the polarization puzzle of $B$ mesons decaying to two vector mesons.
  
\end{abstract}

\vspace*{2.0cm}

\begin{center}
  Published in 
  Physical Review Letters 136 (2026) 021803
\end{center}

\vspace{\fill}

{\footnotesize 
\centerline{\copyright~\papercopyright. \href{\paperlicenceurl}{\paperlicence}.}}
\vspace*{2mm}

\end{titlepage}


\newpage
\setcounter{page}{2}
\mbox{~}
%


\renewcommand{\thefootnote}{\arabic{footnote}}
\setcounter{footnote}{0}


\cleardoublepage


\pagestyle{plain} 
\setcounter{page}{1}
\pagenumbering{arabic}


Decays of beauty mesons provide a rich phenomenology to study the Standard Model~(SM) of particle physics and investigate potential new physics effects. 
Charmless $B$-meson decays into two vector mesons ($V$), $\BtoVV$, serve as a powerful probe for the dynamics of weak and strong interactions through their angular distributions.
Since $\BtoVV$ decays generally proceed through transitions at both leading-order~(tree-level) and higher-order~(loop),
they carry information about the SM flavor structure and the Cabibbo--Kobayashi--Maskawa (CKM) matrix~\cite{Lu:2008gi}. 
The interference between the loop and tree-level processes gives rise to the combined charge conjugation and parity ($\CP$) violation, a phenomenon required to explain the matter-antimatter asymmetry of the universe.

Every $\BtoVV$ decay receives contributions from three helicity amplitudes, which, with an appropriate choice of basis, can be described by longitudinal ($\Azero$), perpendicular ($\Aperp$), and parallel ($\Apara$) components encoding different orbital angular momentum configurations between the two vector mesons~\cite{Beneke:2006hg}, allowing to study the spin structure of the flavor-changing interaction.
These amplitudes can be distinguished by studying the angular distributions of the decay products of the two vector mesons, enabling measurements of the weak and strong phases, as well as the longitudinal polarization fraction  \mbox{$f_L\equiv|\Azero|^2/(|\Azero|^2+|\Aperp|^2+|\Apara|^2)$}.
While the SM predicts an almost total longitudinal polarization, $f_L\approx1$, by a naive factorization ansatz~\cite{Korner:1979ci}, $f_L$ is measured to
span the range from around 10\% to almost 100\% across 
different $\BtoVV$ decay modes~\cite{Belle:2003lsm,Belle:2004uca,Belle:2005lvd,Belle:2008vwd,Belle:2013vat,Belle:2015yjj,Belle:2015xfb,LHCb-PAPER-2013-012,LHCb-PAPER-2014-005,LHCb-PAPER-2015-006,LHCb-PAPER-2017-048,LHCb-PAPER-2018-042,LHCb-PAPER-2019-004,LHCb-PAPER-2019-019}.
Explanations proposed to resolve this ``polarization puzzle'' include enhanced contributions from 
weak-annihilation amplitudes~\cite{Li:2004mp,Kagan:2004uw,Huang:2005if,Su:2010vt,Wang:2017hxe}, 
charm-loop diagrams~\cite{Bauer:2004tj,Wang:2017rmh}, 
higher-order corrections~\cite{Zou:2015iwa,Yan:2018fif}, 
final-state interactions~\cite{Colangelo:2004rd,Li:2004ti,Cheng:2004ru,Ladisa:2004bp}, 
as well as physics beyond the SM~\cite{Yang:2004pm,Baek:2005jk,Huang:2005qb,Bao:2008hd,Kim:2007eea,Datta:2003mj}.   
These contributions may also have a significant impact on predictions of \CP asymmetries in \BtoVV decays~\cite{Yan:2018fif}.
To date, a coherent description of \CP violation and polarization in \BtoVV decays remains challenging.

The specific class of $\BtoVV$ decays with \mbox{$\B\to\rho\Kstar$} has been widely studied, 
where, throughout this Letter, $\rho$ and \Kstar represent the $\rho(770)$ and $\Kstar(892)$ mesons, respectively, unless otherwise stated.
The longitudinal fractions are measured to be substantially different depending on the final state, with
$f_L(\ButoRhoKst)=$ \mbox{$0.78\pm0.12$}~\cite{BaBar:2010evf},  \mbox{$f_L(\Bp\to\rhop\Kstarz)=0.48\pm0.08$}~\cite{PDG2024,BaBar:2006afl,Belle:2004uca}, 
$f_L(\Bz\to\rhoz\Kstarz)=$ \mbox{$0.173 \pm0.026$}~\cite{PDG2024,BaBar:2011ryf,LHCb-PAPER-2018-042}, and $f_L(\Bz\to\rhom\Kstarp)=0.38\pm0.13$~\cite{BaBar:2011ryf},
where charge conjugation is implied throughout this Letter unless otherwise specifically stated. 
These observed discrepancies reflect the varying interference patterns between tree-level and loop  processes~\cite{Cheng:2008gxa}, which would also be revealed by measuring the \CP asymmetries. Among these decays, \CP violation has only been established in the longitudinal component of the  $\Bz\to\myRho\Kstarz$ decay~\cite{LHCb-PAPER-2018-042}.
Detailed and precise measurements of $\B\to\rho\Kstar$ decays, compared with various predictions~\cite{Beneke:2006hg,Huang:2005if,Su:2010vt,Wang:2017hxe,Wang:2017rmh,Zou:2015iwa,Cheng:2008gxa,Cheng:2009cn,Ali:1998gb,Li:2021qiw,Chai:2022ptk,Yan:2025ocu},  permit more precise extractions of nonperturbative parameters through global fits~\cite{Yan:2025ocu} and help to shed light on the polarization puzzle and the dynamics of \CP violation in these decays.
They also comprise fundamental input for theoretically clean, model-independent isospin sum rules that are highly sensitive to physics beyond the SM~\cite{Gronau:2005kz}.

In this Letter the measurements of polarization and \CP parameters in the \mbox{$\ButoRhoKst$} decay are reported.
The reconstruction of the $\myRho$ and $\myKst$ mesons are performed in the decays $\myRho\to\pip\pim$ and  $\myKst\to\KS\pip$, respectively.
An amplitude analysis of the $\Bp\to(\pip\pim)(\KS\pip)$ decay  is performed, using $pp$ collision data collected with the \lhcb detector at center-of-mass energies of $\sqrt{s}=7$, $8$ (Run 1 period in years 2011 and 2012) and $13$\tev (Run 2 period in years 2015--2018), corresponding to a total integrated luminosity of $9\invfb$.

The LHCb detector is a single-arm forward spectrometer dedicated to the study of heavy flavor hadrons and is described in detail in Refs.~\cite{LHCb-DP-2008-001,LHCb-DP-2014-002}. 
The online event selection is performed by a trigger~\cite{LHCb-DP-2012-004,LHCb-DP-2019-001}, which consists of a hardware stage followed by a software stage. The hardware trigger selects hadron, muon, electron and photon candidates with high transverse momentum ($\pt$) based on information from the calorimeter and muon systems.
The software stage, which applies a full event reconstruction, identifies a secondary vertex with a significant displacement from any primary $pp$ interaction vertex~(PV) by means of a multivariate algorithm.
Simulated $\ButoRhoKst$ decays are produced with the software described in Refs.~\cite{Sjostrand:2007gs,Sjostrand:2006za,LHCb-PROC-2010-056,Agostinelli:2002hh,Allison:2006ve,LHCb-PROC-2011-006,davidson2015photos,Lange:2001uf}, and are used to model the effects of the detector acceptance, resolution and selection requirements. The simulation is subjected to the same selection requirements applied to data.

In the offline analysis, five charged tracks identified as pions are used to build the $\Bp$ candidate according to the decay sequence \mbox{$\Bu\to\myRho(\to\pip\pim)\myKst(\to\KS\pip)$}, with $\KS\to\pip\pim$.
The pions are required to be inconsistent with being produced from any PV.
The $\KS$, $\myRho$, $\myKst$ and $\Bp$ candidates are required to have well-reconstructed decay vertices~\cite{LHCb-DP-2014-002} and  significant displacement from any PV.
The reconstructed $\KS$ mass should be  consistent with its known value~\cite{PDG2024}.
The masses of $\myRho$ and $\myKst$ candidates are required to be within the ranges $0.30<m_{\pip\pim}<1.10\gevcc$ and  $0.75<m_{\KS\pip}<1.20\gevcc$, respectively, to contain most of the resonant contributions given their large natural widths.
After applying the two mass-window requirements,  candidates with the interchange of the two same-charge pions are negligible.
Any $\Bp$ candidate with mass in the range $4.90<m_{\pip\pim\KS\pip}<5.80\gevcc$ is retained for further analysis.

The $\KS\to\pip\pim$ candidates are contaminated by $\Lz\to\proton\pim$ decays, where the proton is misidentified as a pion. This background is reduced to a negligible level by vetoing $\KS$ candidates with mass consistent with the known \Lz mass~\cite{PDG2024}, when the energy of a pion is calculated under the proton mass hypothesis.
The same approach is applied to suppress decays involving intermediate charmed states. For example, the $\Bp\to\Dzb(\to\KS\pip\pim)\pip$ decay is removed by vetoing candidates with  $\KS\pip\pim$  mass near the known $\Dz$ mass~\cite{PDG2024}.

To further suppress background due to random combinations of final-state tracks, a boosted decision tree (BDT) classifier~\cite{Breiman,AdaBoost} implemented in the TMVA toolkit~\cite{Hocker:2007ht,TMVA4} is employed.
The  variables discriminating signal and background include the geometrical and kinematical properties of the decay.
The BDT classifier is trained using simulated $\ButoRhoKst$ decays as a proxy for the signal and data candidates in the high $\pip\pim\KS\pip$ mass region, $5.40<m_{\pip\pim\KS\pip}<5.80\gevcc$, for the background.
The requirement on the BDT response is optimized simultaneously with the particle identification (PID) of pions, by maximizing the figure-of-merit defined as $S/\sqrt{S+B}$.
The quantities $S$ and $B$ are the signal and background yields estimated in the signal region chosen as $\pm0.043\gevcc$ around the known $\Bp$ mass of $5.279\gevcc$~\cite{PDG2024}, approximately $2.5$ times the experimental mass resolution.

\begin{figure}
    \centering
    \includegraphics[width=0.45\linewidth]{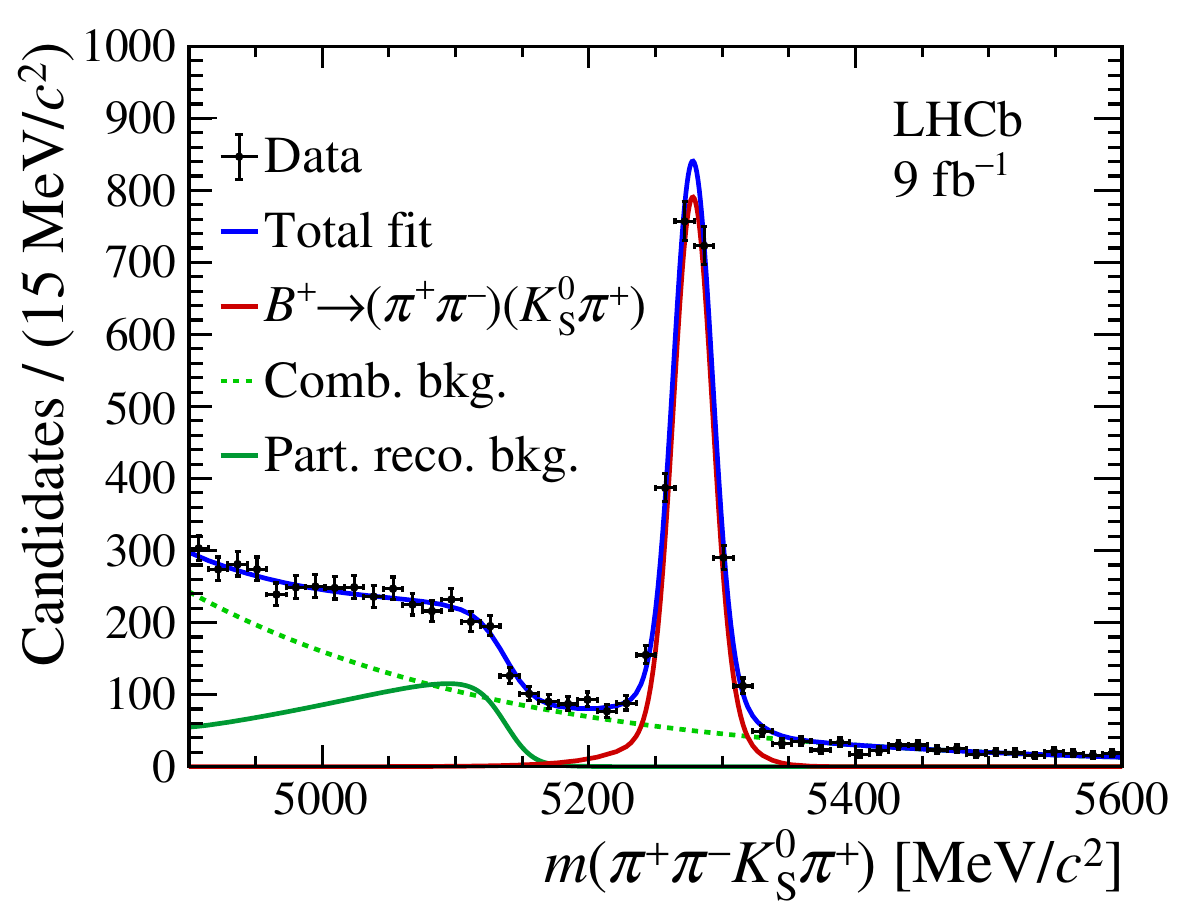}
    \includegraphics[width=0.45\linewidth]{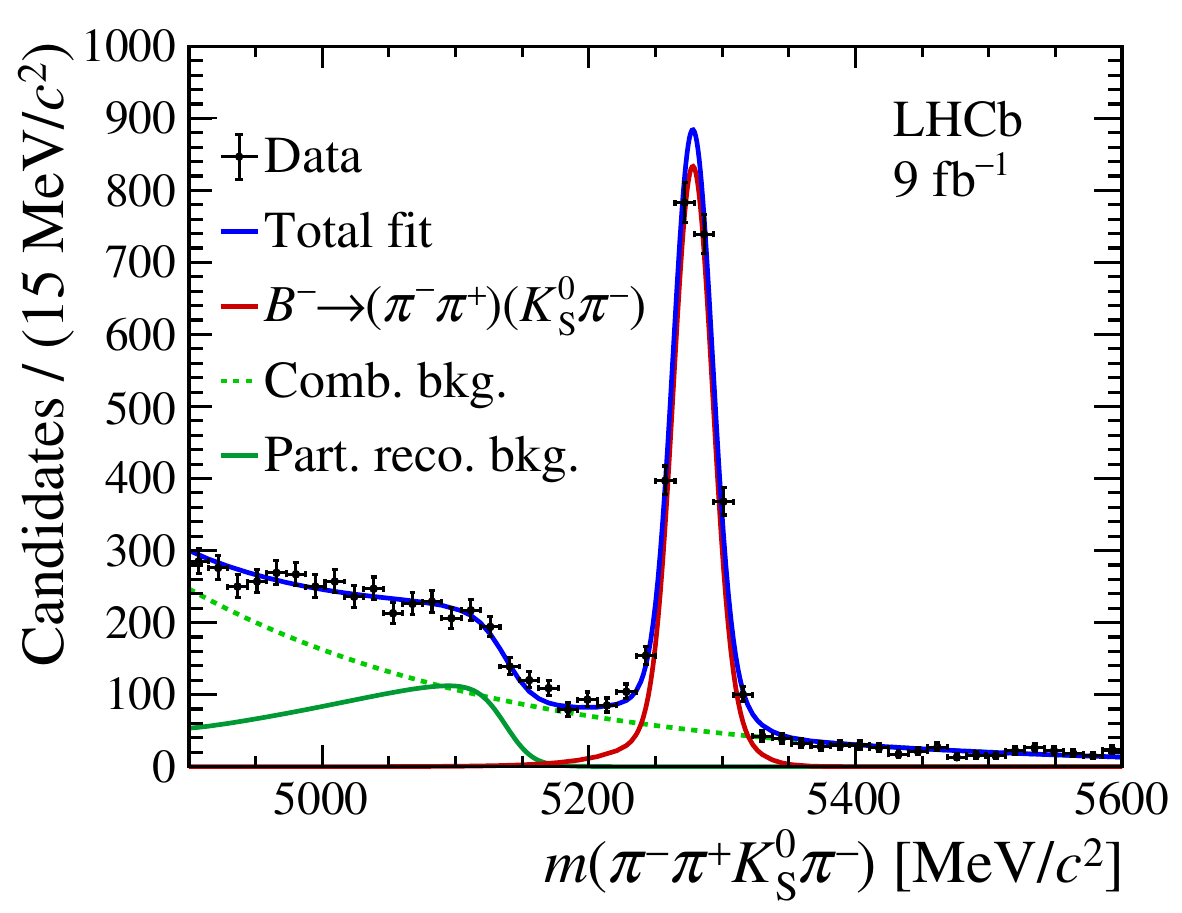}
    \caption{Mass distributions of (left) $B^+\to(\pi^+\pi^-)(K^0_{\mathrm{S}}\pi^+)$ and (right) $B^-\to(\pi^-\pi^+)(K^0_{\mathrm{S}}\pi^-)$ candidates, with the fit results also shown.}
    \label{fig:massfit}
\end{figure}

An extended unbinned maximum-likelihood fit is performed simultaneously to the $m_{\pip\pim\KS\pip}$ and $m_{\pim\pip\KS\pim}$ distributions in the range $4.90<m_{\pipm\pimp\KS\pipm}<5.60\gevcc$, as shown in Fig.~\ref{fig:massfit},
where the shape parameters for each component are shared for \Bp and \Bm decays, with the yields allowed to differ between the two samples.
The fit model consists of the combination of a Gaussian function and a Crystal Ball (CB) function~\cite{Skwarnicki:1986xj} to describe the signal, and an exponential function for the combinatorial background. An ARGUS function~\cite{ARGUS:1990hfq} is used to describe partially reconstructed backgrounds, such as the $B^{0,+}\to (\pip\pim)(\KS\pip)\pi^{-,0}$ decay where the additional $\pi^{-,0}$ meson is not reconstructed.
The numbers of $\Bp$ and $\Bm$ signal candidates are determined to be $2208\pm53$ and $2333\pm55$, respectively, where the uncertainties are statistical. 
The subsequent amplitude analysis exploits candidates in the signal region,
chosen as $5.24<m_{\pipm\pimp\KS\pipm}<5.32\gevcc$.
Within this region, the fraction of \Bp (\Bm) combinatorial background over the total signal and background yield of both flavors is $r_\text{b}^+=0.0631\pm0.0032$ ($r_\text{b}^-=0.0633\pm0.0032$), where the uncertainties are statistical only. 
The charge asymmetry is not significant for either the combinatorial background or the signal.
The amount of partially reconstructed background in the signal regions is negligible.

An amplitude analysis is performed to study the different contributions to the decay, thereby enabling measurements of the polarization and the \CP asymmetry for each component.
The kinematics of the $\Bp\to(\pip\pim)(\KS\pip)$ decay can be fully described by five independent variables, $\vec{\mathcal{O}}=(m_{\pip\pim}, m_{\KS\pip}, \cos \theta_{\pip\pim}, \cos \theta_{\KS\pip}, \phi)$, where $m_{\pip\pim}$ and $m_{\KS\pip}$ are the masses, $\theta_{\pip\pim}$ and $\theta_{\KS\pip}$ are the helicity angles, 
and $\phi$ is the angle between the $\myRho\to\pip\pim$ and $\myKst\to\KS\pip$ decay planes.
The variable $\theta_{\pip\pim}$ ($\theta_{\KS\pip}$) is defined as the angle between the momentum of the $\pip$ ($\KS$) in the $\pip\pim$ ($\KS\pip$) rest frame and the momentum of the $\myRho$ ($\myKst$) meson in the $\Bp$ rest frame.
Following the isobar approach~\cite{Herndon:1973yn}, the total amplitude of the $\Bp\to(\pip\pim)(\KS\pip)$ consists of the coherent sum of quasi-two-body amplitudes.
The differential decay rate for the $\Bp$ decay is defined as the squared modulus of the total amplitude
\begin{equation}
\begin{aligned}
    & \frac{\deriv^5 \Gamma}{
    \deriv\bigl(m_{\pip\pim}\bigr) 
    \deriv\bigl(m_{\KS\pip}\bigr) 
    \deriv\bigl(\cos \theta_{\pip\pim}\bigr) 
    \deriv\bigl(\cos \theta_{\KS\pip}\bigr) 
    \deriv \phi}  \\
    \propto\, & \Phi
    \times\left|\sum_{i=1}^N A_i \times g_i\bigl(\cos \theta_{\pip\pim}, \cos \theta_{\KS\pip}, \phi\bigr) \times M_i\bigl(m_{\pip\pim}, m_{\KS\pip}\bigr)\right|^2,
\label{eq:decay_rate}
\end{aligned}
\end{equation}
where $\Phi$ is the four-body phase-space density
and $A_i$ is the complex coupling for each amplitude.
By construction, the interference terms between components with different spin–parity quantum numbers cancel when integrating over the phase space.
For the $\Bm$ decay, the sign of the $\phi$ angle is flipped and an independent set of complex couplings, $\offsetoverline{A}_i$, is measured.
The $g_i$ functions are characterized by spherical harmonics describing the angular distribution. 
The twenty-two amplitudes included in the baseline fit are detailed in the Supplemental Material~\cite{supplemental}.
The $M_i$ functions describe the mass distributions of the $\pip\pim$ and $\KS\pip$ systems.
They are briefly described below and detailed in the Supplemental Material~\cite{supplemental}.

In the $\pip\pim$ mass spectrum, besides the $\rho(770)^0$ meson,  a few other resonances are identified, including the $\omega(782)$, $f_0(500)$, $f_0(980)$ and $f_0(1370)$ mesons.
The mass distributions of the vector resonances, $\rho(770)^0$ and $\omega(782)$,  are modeled by a Gounaris–Sakurai parameterization~\cite{Gounaris:1968mw} and a relativistic Breit–Wigner function, respectively.
Those for the scalar resonances, $f_0(500)$ and $f_0(1370)$, are described by relativistic Breit–Wigner functions, while that for the $f_0(980)$ meson is modeled by a Flatt\'{e} parameterization~\cite{Flatte:1976xu,Flatte:1976xv,Bugg:2008ig}.
Several fits were performed with and without the scalar $f_0$ states; all states included in the baseline fit have an individual significance above six standard deviations. Alternative descriptions of the $\pi\pi$ S-wave using the K-matrix formalism are considered as a source of systematic uncertainty, which is detailed below.
The $\KS\pip$ mass distribution is dominated by the $\Kstar(892)^+$ resonance and a scalar contribution, referred to as the $(\KS\pip)_\text{S}$ component. 
The $\Kstar(892)^+$ resonance is described by a relativistic Breit–Wigner.
The $(\KS\pip)_\text{S}$ distribution is parameterized by the LASS~\cite{Aston:1987ir} model, which includes the contributions of the $K^*_0(1430)^+$ resonance and a broad low-mass component.
Possible $(\KS\pip)_\text{P}$ contributions from the $K_1(1270)$, $K_1(1400)$ and $K_1^*(1410)$ states were considered, but no significant effects are expected in the studied mass range.
A small fraction of $\Bp\to a_1(1260)^+ \KS$ and $\Bp \to a_1(1640)^+ \KS$ decays is observed in the sample, with $a_1^+\to\myRho(\to\pip\pim)\pip$, $a_1^+\to f_0(500)(\to\pip\pim)\pip$, and $a_1^+\to f_0(980)(\to\pip\pim)\pip$.
The $a_1$ mass distributions are described by relativistic Breit–Wigner functions, with the angular variables redefined accordingly. 
Parameters describing the mass distributions are fixed in the baseline fit according to the results of previous measurements~\cite{PDG2024,LHCb-PAPER-2018-042}.

\begin{figure}[!tb]
    \centering
    \includegraphics[width=1.0\linewidth]{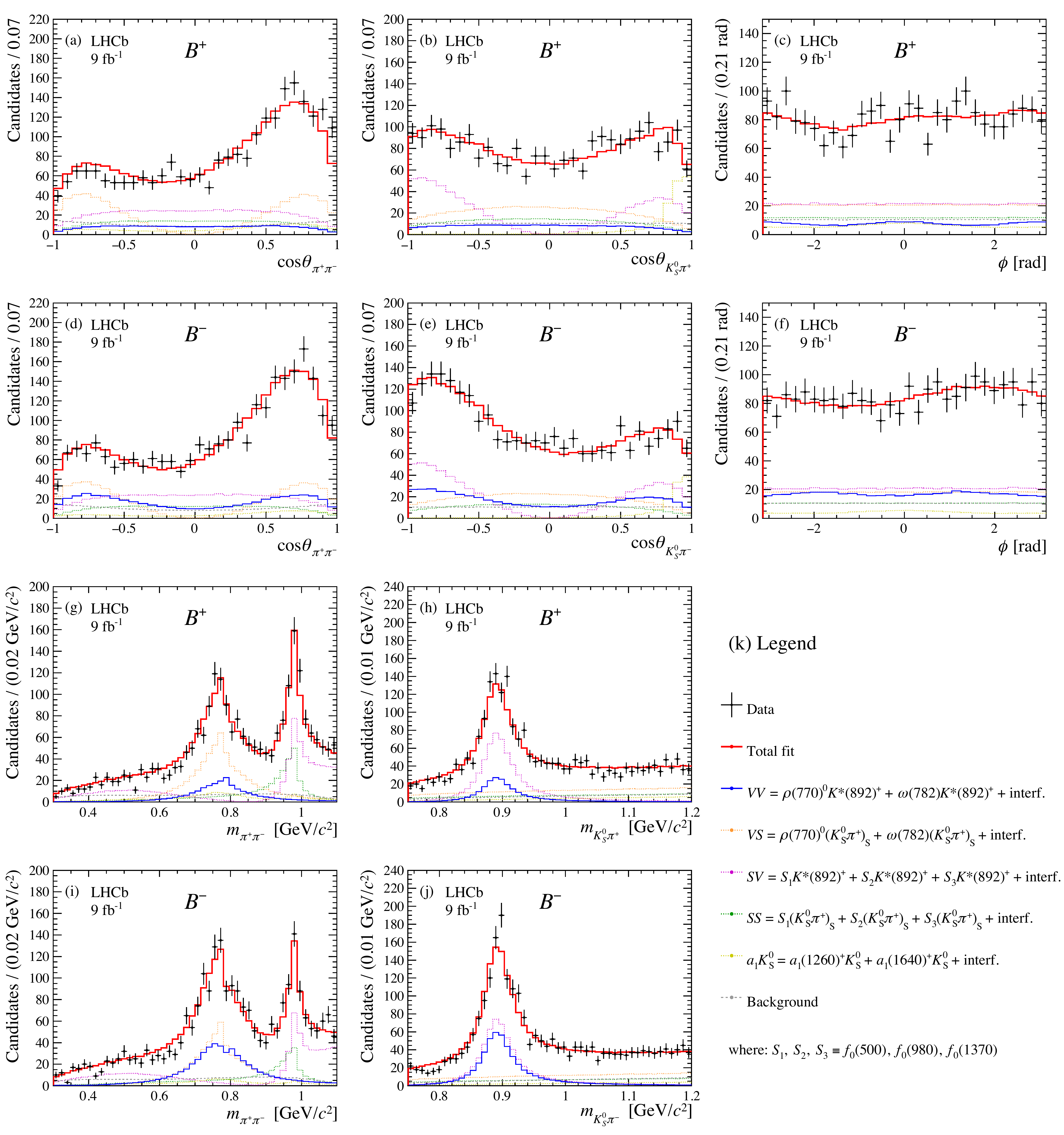}
    \caption{Projections of the multidimensional amplitude fit onto (a) $\cos \theta_{\pi^+\pi^-}$, (b) $\cos \theta_{K^0_{\mathrm{S}}\pi^+}$, (c) $\phi$, (g) $m_{\pi^+\pi^-}$, and (h) $m_{K^0_{\mathrm{S}}\pi^+}$ for the $B^+\to(\pi^+\pi^-)(K^0_{\mathrm{S}}\pi^+)$ decay.
    Plots (d)(e)(f)(i)(j) show the corresponding variables for the $B^-\to(\pi^-\pi^+)(K^0_{\mathrm{S}}\pi^-)$ decay.
    }
    \label{fig:AmAn}
\end{figure}

The combined probability density function (PDF) for $\Bp$ and $\Bm$ decays, composed of signal and background components, is given by
\begin{equation}
\begin{aligned}
  \mathcal{P}_{\text {tot }}(\vec{\mathcal{O}},q; \{A_i,\offsetoverline{A}_i\})=&\frac{(1+q) \mathcal{P}_\text{s}+(1-q) \offsetoverline{\mathcal{P}}_\text{s}}{2\int\left(\mathcal{P}_\text{s}+\offsetoverline{\mathcal{P}}_\text{s}\right)\deriv \vec{\mathcal{O}}}\left(1-r_\text{b}^+-r_\text{b}^-\right)\\
&+
\frac{(1+q)\mathcal{P}_\text{b}}{2\int\mathcal{P}_\text{b}\deriv \vec{\mathcal{O}}}r_\text{b}^{+} +
\frac{(1-q)\offsetoverline{\mathcal{P}}_\text{b}}{2\int\offsetoverline{\mathcal{P}}_\text{b}\deriv \vec{\mathcal{O}}}r_\text{b}^{-} \, ,
\end{aligned}
\label{eq:totalPDF}
\end{equation}
where $q=\pm 1$ indicates the charge of the $B^\pm$ meson,
and $r_\text{b}^{\pm}$ are background fractions fixed to the values obtained from the mass fit.
The component $\mathcal{P}_\text{s}$  ($\offsetoverline{\mathcal{P}}_\text{s}$) is the differential decay rate described in Eq.~\eqref{eq:decay_rate} multiplied by the experimental efficiency for the $\Bp$~($\Bm$) decay.  
The efficiency map as a function of the five kinematic variables is obtained by parameterizing the simulation sample using five-dimensional Legendre polynomials.
Data-driven corrections are applied to the simulation to account for imperfections in the detector response and asymmetries in the detection efficiency of $\pip$ and $\pim$ tracks. 
The background distributions, $\mathcal{P}_\text{b}$ and  $\offsetoverline{\mathcal{P}}_\text{b}$, for $\Bp$ and $\Bm$ candidates respectively, are parameterized by five-dimensional Legendre polynomials using candidates in the high-mass region of the $\Bpm$ spectra. 
The amplitude fit is performed simultaneously to the \mbox{Run 1} and \mbox{Run 2} samples.
In the fit, the phase of each amplitude is measured relative to that of the $\Bp\to\myRho(\KS\pip)_\text{S}$ component, which has a relatively large contribution.
As the phase difference between \Bp and \Bm is not observable, the phase of the $\Bm\to\myRho(\KS\pim)_\text{S}$ component is set to zero.
The one-dimensional projections of the amplitude fit in the five kinematic variables are shown in 
Fig.~\ref{fig:AmAn}.
The \CP asymmetry between the couplings of the $\Bp\to\myRho(\KS\pip)_\text{S}$ amplitude and its charge conjugate is found to be consistent with zero.

Two main categories of systematic uncertainty contribute to the result: those originating from experimental assumptions and those related to the amplitude model. 
The experimental category includes effects due to the limited precision of the background fractions, the efficiency map and background modeling.
The amplitude  model uncertainties arise from limited knowledge of the fixed parameters of the resonance mass models, and possible spin-2 components neglected in the baseline fit.
Systematic uncertainties due to experimental effects are found to be subdominant. 
The breakdown of the systematic uncertainties of observables is detailed in Table~\ref{tab:results} of the End Matter.
Results of the fitted parameters with their uncertainties are also summarized in Table~\ref{tab:fit-results} of the End Matter.

The experimental systematic uncertainty from the background fractions is evaluated by varying the $r_b^\pm$ parameters within their uncertainties and repeating the amplitude fit. 
Systematic uncertainties due to the efficiency parameterization arise from imperfections of the five-dimensional efficiency map and from any residual charge-asymmetry effect. 
Imperfections are corrected by a weighting procedure to improve agreement with simulation across the five-dimensional space.
The charge-asymmetry effects are assessed by deriving separate efficiency models for $\Bp$ and $\Bm$ decays.
To evaluate the uncertainty due to the background PDF, an alternative parametrization is obtained by combining $\Bp$ and $\Bm$ samples or using samples with more relaxed selection criteria.
Also, pseudoexperiments by bootstrapping the background sample~\cite{efron:1979} are used to extract the background PDF, where for each pseudoexperiment the amplitude fit is performed again.
In each case, the deviation from the baseline result or the standard deviation of the results among the pseudoexperiments is taken as the corresponding systematic uncertainty.

The systematic uncertainty due to
fixed parameters in the amplitude model is estimated using repeated fits to the data.
Each time the masses, widths, and hadron radii of intermediate resonances are varied within their uncertainties and an amplitude fit is performed under this model.
The standard deviation of the distribution of the fit results among the repeated fits is taken as the corresponding systematic uncertainty.
For the $B\to VV$ decay, the relative orbital angular momentum of the two vector mesons, $L$,  could take the values $L=0$ or $L=2$ for the $\Azero$ and $\Apara$ amplitudes, and $L=1$ for the $\Aperp$ amplitude. Various $L$ configurations are studied and the results are compared with that of the baseline fit with $L=0$, and the largest difference is taken as the systematic uncertainty.
The scalar component of the $\pip\pim$ mass spectrum can also be modeled using the K-matrix approach~\cite{Chung:1995dx}, as detailed in~\cite{Back:2017zqt}.
The difference between the results obtained with this alternative model and those from the baseline fit using explicit $f_0$ resonances is taken as a systematic uncertainty.
Furthermore, possible contributions of spin-2 resonances to the decay are assessed by including the $f_2(1270)\to\pip\pim$ or $K_2^*(1430)^+\to\KS\pip$ components in the fit.
The systematic uncertainties are estimated using the mean value of deviations, between the fits using extended model and baseline model, over an ensemble of pseudoexperiments.

The longitudinal polarization fraction  averaged over the $\ButoRhoKst$ and \mbox{$\Bm\to\rhoz\Kstarm$} final states is determined in terms of the $\AiVV$ and $\AiVVbar$ amplitudes, with $\lambda\in\{0,\parallel,\perp\}$, to be
\begin{equation*}
        f_L \equiv\frac{|\AzeroVV|^2+|\AzeroVVbar|^2}{\sum_{\lambda}(|\AiVV|^2+|\AiVVbar|^2)}= \fLResultSimple,
    \label{eq:fL_def}
\end{equation*}
where, throughout this Letter,  the first uncertainty is statistical and the second is systematic. 
The result is consistent with and more precise than the previous measurement~\cite{BaBar:2010evf}.
The charge-specific quantities are determined to be
\begin{equation*}
    f_L^+ \equiv\frac{|\AzeroVV|^2}{\sum_{\lambda}|\AiVV|^2}= \fLpResultSimple,
    \quad
    f_L^- \equiv\frac{|\AzeroVVbar|^2}{\sum_{\lambda}|\AiVVbar|^2}= \fLmResultSimple.
\end{equation*}

The direct \CP asymmetry for the total $\ButoRhoKst$ decay rates~\cite{Wang:2017rmh,Yan:2025ocu} is measured to be
\begin{equation*}
    \mathcal{A}_{\CP}\equiv
    \frac{\sum_{\lambda}(|\AiVVbar|^2-|\AiVV|^2)}{\sum_{\lambda}(|\AiVVbar|^2+|\AiVV|^2)}
    =\ARhoKstResultSimple.
\end{equation*}
The \CP asymmetries are also determined for the magnitude and phase of each $\ButoRhoKst$ amplitude. The results for the longitudinal component are
 \begin{equation*}
     \begin{aligned}
         \mathcal{A}_{\CP}(\AzeroVV) &\equiv\frac{|\AzeroVVbar|^2-|\AzeroVV|^2}{|\AzeroVVbar|^2+|\AzeroVV|^2}&= \ARhoKstLResultSimple,\phantom{\mathrm{rad}}\\
         \Delta_{\CP}(\AzeroVV)  &\equiv \arg(\AzeroVVbar)-\arg(\AzeroVV)&= \AphaRhoKstLResultSimple\,\mathrm{rad},
     \end{aligned}
 \end{equation*}
while the \CP asymmetries of squared magnitudes and phases of \AparaVV and \AperpVV components are measured to be
\begin{equation*}
    \begin{aligned}
        \mathcal{A}_{\CP}(\AparaVV) &= \ARhoKstParaResultSimple,\phantom{{\mathrm{rad}}} \\
        \Delta_{\CP}(\AparaVV) &=\phantom{-} \AphaRhoKstParaResultSimple\,\mathrm{rad},\\
        \mathcal{A}_{\CP}(\AperpVV) &=\phantom{-} \ARhoKstPerpResultSimple,\phantom{\mathrm{rad}} \\
        \Delta_{\CP}(\AperpVV) &=\phantom{-} \AphaRhoKstPerpResultSimple\,\mathrm{rad}.\\
    \end{aligned}
    \label{eq:para-perp-results}
\end{equation*}
Specifically, sizeable \CP violation is observed for the magnitude of the longitudinal component, which drives the \CP violation of the overall $\ButoRhoKst$ decay rate.  
The phase difference between the longitudinal component of the $\ButoRhoKst$ decay and its charge conjugate is 3.9 standard deviations away from zero,
which suggests that long-distance final-state interactions with other contributions in the phase space are playing a role in \CP violation.
The significance of \CP violation for the $\ButoRhoKst$ decay is quantified using a likelihood-ratio test~\cite{Wilks:1938dza} between the baseline fit, which allows for \CP violation, and a \CP-conserving model where the couplings of the $\ButoRhoKst$ decay and its charge conjugate are constrained to be identical. The significance of a nonvanishing \CP asymmetry is found to be above nine standard deviations with systematic uncertainties taken into account. It marks the first observation of \CP violation in this decay.

Triple Product Asymmetries (TPAs), which exploit the interference between two amplitudes, have been widely used to study \CP violation for multibody decays.
They are defined to be odd under both the time reversal and parity transformations.
A nonzero TPA can either be due to a \CP-violating phase or a \CP-conserving phase induced by final-state interactions, referred to as the
$\it{true}$ and $\it{fake}$ TPA, respectively.
The $\it{true}$ and $\it{fake}$ TPAs are related to the couplings of the amplitudes~\cite{Gronau:2011cf,Li:2021qiw} as
\begin{equation}
\begin{aligned}
\mathcal{A}_{\substack{\it{true}\\\it{fake}}}^{(1)} 
& \equiv-\frac{2 \sqrt{2}}{\pi} \frac{\Imag(\AperpVV \AzeroVV^*\mp\AperpVVbar 
\AzeroVVbarbis)}{\sum_{\lambda\in\{0,\parallel,\perp\}}(|\AiVV|^2+|\AiVVbar|^2)}, \\
\mathcal{A}_{\substack{\it{true}\\\it{fake}}}^{(2) } 
& \equiv-\frac{4}{\pi} \frac{\Imag(\AperpVV \AparaVV^*\mp\AperpVVbar 
\AparaVVbarbis)}{\sum_{\lambda\in\{0,\parallel,\perp\}}(|\AiVV|^2+|\AiVVbar|^2)},
\end{aligned}
\label{eq:TPA_VV_chg-avg}
\end{equation}
where $\AiVV^*$ ($\AiVVbarbis$) denotes the complex conjugate of $\AiVV$ ($\AiVVbar$).
Using the amplitude fit results, the TPAs are computed to be
\begin{equation*}
    \begin{aligned}
    \mathcal{A}_{\it{true}}^{(1)} &= -0.105\pm0.024\pm0.013,\\
    \mathcal{A}_{\it{true}}^{(2)} &= \phantom{-}0.007\pm0.019\pm0.003,\\
    \mathcal{A}_{\it{fake}}^{(1)} &= -0.157\pm0.024\pm0.011,\\
    \mathcal{A}_{\it{fake}}^{(2)} &= \phantom{-}0.008\pm0.018\pm0.004.
    \end{aligned}
    \label{eq:TPA-results}
\end{equation*}
The $\it{true}$ TPA $\mathcal{A}_{\it{true}}^{(1) }$ deviates from zero with a significance of approximately four standard deviations, indicating evidence of \CP violation in interference. The $\it{fake}$ TPA $\mathcal{A}_{\it{fake}}^{(1) }$ is nonzero with a significance of about six standard deviations, marking parity violation in interference~\cite{Datta:2011qz}.

In summary, an amplitude analysis of the decay \mbox{$\Bp\to(\pip\pim)(\KS\pip)$} is performed within the mass ranges \mbox{$0.30<m_{\pip\pim}<1.10\gevcc$} and \mbox{$0.75<m_{\KS\pip}<1.20\gevcc$}.
The longitudinal polarization and \CP asymmetries are studied for the \mbox{$\Bp\to\rho(770)^0\Kstar(892)^+$} decay.
The  \CP-averaged longitudinal fraction is measured to be \mbox{$f_L = \fLResult$}.
The direct \CP asymmetry is determined to be $\mathcal{A}_{\CP} = \ARhoKstResult$, and the significance of \CP violation is measured to  exceed nine standard deviations using a likelihood-ratio test. 
This is the first observation of \CP violation in the \mbox{$\Bp\to\rho(770)^0\Kstar(892)^+$} decay.
These measurements are consistent with those previously reported by the \babar collaboration~\cite{BaBar:2010evf}, with much better precision.
The \CP asymmetries are also derived for the magnitudes and phases of the three subamplitudes, suggesting that the longitudinal component drives the total \CP violation.
These results help to constrain theoretical models used to calculate $B$-hadron decays, shed light on the dynamics of heavy-flavor decays and contribute to the understanding of the polarization puzzle of \B-meson decays.
\section*{Acknowledgements}
\noindent We express our gratitude to our colleagues in the CERN
accelerator departments for the excellent performance of the LHC. We
thank the technical and administrative staff at the LHCb
institutes.
We acknowledge support from CERN and from the national agencies:
ARC (Australia);
CAPES, CNPq, FAPERJ and FINEP (Brazil); 
MOST and NSFC (China); 
CNRS/IN2P3 (France); 
BMBF, DFG and MPG (Germany); 
INFN (Italy); 
NWO (Netherlands); 
MNiSW and NCN (Poland); 
MCID/IFA (Romania); 
MICIU and AEI (Spain);
SNSF and SER (Switzerland); 
NASU (Ukraine); 
STFC (United Kingdom); 
DOE NP and NSF (USA).
We acknowledge the computing resources that are provided by ARDC (Australia), 
CBPF (Brazil),
CERN, 
IHEP and LZU (China),
IN2P3 (France), 
KIT and DESY (Germany), 
INFN (Italy), 
SURF (Netherlands),
Polish WLCG (Poland),
IFIN-HH (Romania), 
PIC (Spain), CSCS (Switzerland), 
and GridPP (United Kingdom).
We are indebted to the communities behind the multiple open-source
software packages on which we depend.
Individual groups or members have received support from
Key Research Program of Frontier Sciences of CAS, CAS PIFI, CAS CCEPP, 
Fundamental Research Funds for the Central Universities,  and Sci.\ \& Tech.\ Program of Guangzhou (China);
Minciencias (Colombia);
EPLANET, Marie Sk\l{}odowska-Curie Actions, ERC and NextGenerationEU (European Union);
A*MIDEX, ANR, IPhU and Labex P2IO, and R\'{e}gion Auvergne-Rh\^{o}ne-Alpes (France);
Alexander-von-Humboldt Foundation (Germany);
ICSC (Italy); 
Severo Ochoa and Mar\'ia de Maeztu Units of Excellence, GVA, XuntaGal, GENCAT, InTalent-Inditex and Prog.~Atracci\'on Talento CM (Spain);
SRC (Sweden);
the Leverhulme Trust, the Royal Society and UKRI (United Kingdom).

\begingroup
  \makeatletter
    \let\addcontentsline\@gobblethree
  \makeatother
  \clearpage

\section*{End Matter}

The breakdown of systematic uncertainty contributions for the reported observables is shown in Table~\ref{tab:results}.

\begin{table}[h]
    \centering
    \caption{Summary of the systematic uncertainty contributions for the reported observables, with statistical and total systematic uncertainties shown as reference. These uncertainties are given as absolute values.}
    \centering
    \scalebox{0.9}{
    \begin{tabular}{lcc|ccccc} \hline\hline
    & \multirow{2}{*}{Stat.} & Syst. & Background & Efficiency & Background & Mass & Spin-2 \\
    & & (total) & fractions & map & modeling & models & components \\
    \hline
    $\mathcal{A}_{\CP}$            & 0.062 & 0.024 & 0.0001 & 0.0049 & 0.0066 & 0.0216 & 0.0067 \\
    $\mathcal{A}_{\CP}(\AzeroVV)$            & 0.083 & 0.029 & 0.0003 & 0.0061 & 0.0114 & 0.0206 & 0.0151 \\
    $\mathcal{A}_{\CP}(\AparaVV)$            & 0.137 & 0.027 & 0.0015 & 0.0040 & 0.0175 & 0.0166 & 0.0116 \\
    $\mathcal{A}_{\CP}(\AperpVV)$            & 0.140 & 0.051 & 0.0005 & 0.0035 & 0.0158 & 0.0467 & 0.0116 \\
    $\Delta_{\CP}(\AzeroVV)$            & 0.177 & 0.048 & 0.0031 & 0.0129 & 0.0304 & 0.0314 & 0.0141 \\
    $\Delta_{\CP}(\AparaVV)$            & 0.187 & 0.114 & 0.0040 & 0.0021 & 0.0138 & 0.1092 & 0.0282 \\
    $\Delta_{\CP}(\AperpVV)$            & 0.180 & 0.122 & 0.0013 & 0.0053 & 0.0175 & 0.1145 & 0.0365 \\
    \hline\hline
    \end{tabular}
    }
    \label{tab:results}
\end{table}

Results of the fitted amplitude parameters, \ie the real and imaginary parts of amplitude couplings, are summarized in Table~\ref{tab:fit-results}.
Asymmetric uncertainties are evaluated following the procedure described in Ref.~\cite{Barlow:2024iwp}.
Parameters marked with ``---'' are constrained in the fit, as explained below.

The reference amplitude is taken to be $A_{\rho(\KS\pi)_\text{S}}$, with couplings fixed to
\begin{equation}
    \begin{split}
        \Bp: \Real(A_{\rho(\KS\pi)_\text{S}}^+) = 1-\Delta, \quad \Imag(A_{\rho(\KS\pi)_\text{S}}^+) = 0, \\
        \Bm: \Real(A_{\rho(\KS\pi)_\text{S}}^-) = 1+\Delta, \quad \Imag(A_{\rho(\KS\pi)_\text{S}}^-) = 0,
    \end{split}
\end{equation}
where $\Delta$ describes the \CP asymmetry of the reference amplitude.
In the fit, $\Real(A_{\rho(\KS\pi)_\text{S}}^+)$ is thus constrained to be $2-\Real(A_{\rho(\KS\pi)_\text{S}}^-)$.

For each $a_1$ resonance, the decays $a_1\to X\pi$ ($X=\rho, f_0(500), f_0(980)$) proceed via \CP-conserving strong interactions.
The \CP-violating contributions are therefore shared among all $B\to a_1\KS$ channels, where $a_1$ denotes $a_1(1260)^+$ or $a_1(1640)^+$.
The four amplitudes satisfy
\begin{equation}
    \frac{A_0^+}{A_0^-}=\frac{A_1^+}{A_1^-}=\frac{A_2^+}{A_2^-}=\frac{A_3^+}{A_3^-},
\end{equation}
with 
\begin{equation}
\begin{split}
    A_0=A_{a_1(\to \rho \pi)\KS}^0&, \quad A_1=A_{a_1(\to \rho \pi)\KS}^\parallel, \\
    \quad A_2=A_{a_1(\to f_0(500) \pi)\KS}&, \quad A_3=A_{a_1(\to f_0(980) \pi)\KS}.
    \end{split}
\end{equation}
In the fit, $A_0^{+}$ and all $A_{i=0,1,2,3}^{-}$ are free parameters, while the other $A_{i=1,2,3}^{+}$ are constrained through these relations.

\begin{table}[b]
    \centering
    \caption{Summary of the fitted real and imaginary parts of the amplitude couplings for the \Bp and \Bm samples, respectively. The first uncertainties are statistical and the second ones are systematic. The reference amplitude is taken to be $A_{\rho(\KS\pi)_\text{S}}$.
    Parameters for the $a_1^+\KS$ components are reduced by constraining the \CP-violating contributions to be the same among decay channels for each $a_1$ resonance, since its subdecays are \CP-conserving strong interactions. These are signified with a ``---'' in the Table.}
    \centering
        \scalebox{0.83}{
        \begin{tabular}{l|rll|rll} 
        \hline\hline
        Parameter & \multicolumn{3}{c|}{$\Bp$} & \multicolumn{3}{c}{$\Bm$} \\
        \hline
        $\Real(A_{\rho(\KS\pi)_\text{S}})$ & \multicolumn{3}{c|}{---} & $0.9735$ & $\pm 0.0314$ & $\pm 0.0142$ \\
        $\Imag(A_{\rho(\KS\pi)_\text{S}})$ && $0$ (Fixed) &&& $0$ (Fixed) &\\
        $\Real(A_{\rho\Kstar}^0)$ & $0.1641$ & $\pm 0.0254$ & $\pm 0.0415$ & $0.2392$ & $\pm 0.0332$ & $\pm 0.0750$ \\
        $\Imag(A_{\rho\Kstar}^0)$ & $0.0242$ & $\pm 0.0266$ & $\pm 0.0316$ & $0.2811$ & $\pm 0.0326$ & $\pm 0.0779$ \\
        $\Real(A_{\rho\Kstar}^\parallel)$ & $-0.1289$ & $\pm 0.0141$ & $\pm 0.0332$ & $-0.1203$ & $\pm 0.0143$ & $\pm 0.0293$ \\
        $\Imag(A_{\rho\Kstar}^\parallel)$ & $0.0297$ & $\pm 0.0184$ & $\pm 0.0307$ & $-0.0308$ & $\pm 0.0180$ & $\pm 0.0209$ \\
        $\Real(A_{\rho\Kstar}^\perp)$ & $-0.1021$ & $\pm 0.0142$ & $\pm 0.0261$ & $-0.1385$ & $\pm 0.0133$ & $\pm 0.0325$ \\
        $\Imag(A_{\rho\Kstar}^\perp)$ & $0.0246$ & $\pm 0.0160$ & $\pm 0.0249$ & $-0.0246$ & $\pm 0.0174$ & $\pm 0.0248$ \\
        $\Real(A_{\omega\Kstar}^0)$ & $0.0005$ & $\pm 0.0067$ & $\pm 0.0059$ & $-0.0087$ & $\pm 0.0056$ & $\pm 0.0034$ \\
        $\Imag(A_{\omega\Kstar}^0)$ & $-0.0114$ & $\pm 0.0058$ & $\pm 0.0126$ & $-0.0119$ & $\pm 0.0074$ & $\pm 0.0029$ \\
        $\Real(A_{\omega\Kstar}^\parallel)$ & $-0.0023$ & $\pm 0.0030$ & $\pm 0.0009$ & $-0.0006$ & $\pm 0.0023$ & $\pm 0.0005$ \\
        $\Imag(A_{\omega\Kstar}^\parallel)$ & $-0.0025$ & $\pm 0.0028$ & $\pm 0.0013$ & $0.0020$ & $\pm 0.0022$ & $\pm 0.0007$ \\
        $\Real(A_{\omega\Kstar}^\perp)$ & $-0.0001$ & $\pm 0.0044$ & $\pm 0.0014$ & $-0.0027$ & $\pm 0.0033$ & $\pm 0.0008$ \\
        $\Imag(A_{\omega\Kstar}^\perp)$ & $-0.0042$ & $\pm 0.0030$ & $\pm 0.0013$ & $0.0015$ & $\pm 0.0034$ & $\pm 0.0007$ \\
        $\Real(A_{\omega(\KS\pi)_\text{S}})$ & $0.0052$ & $\pm 0.0112$ & $\pm 0.0041$ & $-0.0001$ & $\pm 0.0119$ & $\pm 0.0019$ \\
        $\Imag(A_{\omega(\KS\pi)_\text{S}})$ & $0.0254$ & $\pm 0.0096$ & $\pm 0.0048$ & $0.0276$ & $\pm 0.0087$ & $\pm 0.0026$ \\
        $\Real(A_{f_0(500)\Kstar})$ & $1.5539$ & $\pm 0.1852$ & $\pm 0.3636$ & $1.0127$ & $\pm 0.2455$ & $\pm 0.2885$ \\
        $\Imag(A_{f_0(500)\Kstar})$ & $0.3152$ & $\pm 0.3778$ & $\pm 0.2541$ & $1.1808$ & $\pm 0.2386$ & $\pm 0.3634$ \\
        $\Real(A_{f_0(980)\Kstar})$ & $-0.1784$ & ${}_{-0.1102}^{+0.0909}$ & ${}^{+0.1183}_{-0.1341}$ & $-0.4046$ & ${}_{-0.1624}^{+0.0667}$ & ${}^{+0.0913}_{-0.1537}$ \\
        $\Imag(A_{f_0(980)\Kstar})$ & $-0.6713$ & ${}_{-0.3013}^{+0.0811}$ & ${}^{+0.1309}_{-0.1595}$ & $-0.6048$ & ${}_{-0.2850}^{+0.0824}$ & ${}^{+0.1284}_{-0.1519}$ \\
        $\Real(A_{f_0(1370)\Kstar})$ & $0.5053$ & ${}_{-0.3195}^{+0.3206}$ & ${}^{+0.2821}_{-0.2658}$ & $1.0462$ & ${}_{-0.2688}^{+0.1879}$ & ${}^{+0.3300}_{-0.2286}$ \\
        $\Imag(A_{f_0(1370)\Kstar})$ & $0.1557$ & ${}_{-0.2424}^{+0.2860}$ & ${}^{+0.1668}_{-0.1855}$ & $0.4208$ & ${}_{-0.2555}^{+0.2378}$ & ${}^{+0.2246}_{-0.2077}$ \\
        $\Real(A_{f_0(500)(\KS\pi)_\text{S}})$ & $0.6749$ & $\pm 0.4192$ & $\pm 0.2929$ & $1.3057$ & $\pm 0.3611$ & $\pm 0.2671$ \\
        $\Imag(A_{f_0(500)(\KS\pi)_\text{S}})$ & $-0.3747$ & $\pm 0.4989$ & $\pm 0.2675$ & $-0.1689$ & $\pm 0.5528$ & $\pm 0.2425$ \\
        $\Real(A_{f_0(980)(\KS\pi)_\text{S}})$ & $-0.2698$ & ${}_{-0.2642}^{+0.0870}$ & ${}^{+0.1513}_{-0.1596}$ & $-0.3532$ & ${}_{-0.2573}^{+0.0924}$ & ${}^{+0.0725}_{-0.0790}$ \\
        $\Imag(A_{f_0(980)(\KS\pi)_\text{S}})$ & $1.3567$ & ${}_{-0.2222}^{+0.6411}$ & ${}^{+0.3352}_{-0.3304}$ & $0.9319$ & ${}_{-0.1349}^{+0.4909}$ & ${}^{+0.3050}_{-0.3044}$ \\
        $\Real(A_{f_0(1370)(\KS\pi)_\text{S}})$ & $1.3511$ & ${}_{-0.5138}^{+0.5718}$ & ${}^{+0.5271}_{-0.5279}$ & $0.2864$ & ${}_{-0.5300}^{+0.5436}$ & ${}^{+0.4094}_{-0.4008}$ \\
        $\Imag(A_{f_0(1370)(\KS\pi)_\text{S}})$ & $0.1008$ & ${}_{-0.4300}^{+1.0180}$ & ${}^{+0.6857}_{-0.6563}$ & $0.5882$ & ${}_{-0.4542}^{+0.4970}$ & ${}^{+0.3376}_{-0.3491}$ \\
        $\Real(A_{(a_1(1260)\to \rho \pi)\KS}^0)$ & $-2.5802$ & $\pm 1.3487$ & $\pm 2.2671$ & $-2.0785$ & $\pm 0.6787$ & $\pm 0.9635$ \\
        $\Imag(A_{(a_1(1260)\to \rho \pi)\KS}^0)$ & $-0.5676$ & $\pm 0.9027$ & $\pm 1.1999$ & $-0.8797$ & $\pm 0.6419$ & $\pm 0.8860$ \\
        $\Real(A_{(a_1(1260)\to \rho \pi)\KS}^\parallel)$ & \multicolumn{3}{c|}{---} & $-0.1507$ & $\pm 1.5943$ & $\pm 0.8448$ \\
        $\Imag(A_{(a_1(1260)\to \rho \pi)\KS}^\parallel)$ & \multicolumn{3}{c|}{---} & $3.0275$ & $\pm 1.3924$ & $\pm 1.4278$ \\
        $\Real(A_{(a_1(1260)\to f_0(500) \pi)\KS})$ & \multicolumn{3}{c|}{---} & $-8.5584$ & ${}_{-1.8408}^{+1.8330}$ & ${}^{+4.2256}_{-3.9197}$ \\
        $\Imag(A_{(a_1(1260)\to f_0(500) \pi)\KS})$ & \multicolumn{3}{c|}{---} & $3.5230$ & ${}_{-0.7814}^{+1.2376}$ & ${}^{+2.0371}_{-2.0247}$ \\
        $\Real(A_{(a_1(1260)\to f_0(980) \pi)\KS})$ & \multicolumn{3}{c|}{---} & $1.2407$ & ${}_{-1.2627}^{+0.8240}$ & ${}^{+1.3965}_{-1.3885}$ \\
        $\Imag(A_{(a_1(1260)\to f_0(980) \pi)\KS})$ & \multicolumn{3}{c|}{---} & $0.7200$ & ${}_{-1.0314}^{+0.8666}$ & ${}^{+0.4065}_{-0.4259}$ \\
        $\Real(A_{(a_1(1640)\to \rho \pi)\KS}^0)$ & $-0.3555$ & $\pm 0.4472$ & $\pm 0.3954$ & $-0.2564$ & $\pm 0.4432$ & $\pm 0.3951$ \\
        $\Imag(A_{(a_1(1640)\to \rho \pi)\KS}^0)$ & $-0.0363$ & $\pm 0.5007$ & $\pm 0.3456$ & $-0.0859$ & $\pm 0.3758$ & $\pm 0.2064$ \\
        $\Real(A_{(a_1(1640)\to \rho \pi)\KS}^\parallel)$ & \multicolumn{3}{c|}{---} & $-0.3892$ & $\pm 0.8504$ & $\pm 0.5404$ \\
        $\Imag(A_{(a_1(1640)\to \rho \pi)\KS}^\parallel)$ & \multicolumn{3}{c|}{---} & $-0.2174$ & $\pm 0.9290$ & $\pm 0.9068$ \\
        $\Real(A_{(a_1(1640)\to f_0(500) \pi)\KS})$ &  \multicolumn{3}{c|}{---} & $3.8584$ & $\pm 1.3145$ & $\pm 1.4161$ \\
        $\Imag(A_{(a_1(1640)\to f_0(500) \pi)\KS})$ &  \multicolumn{3}{c|}{---} & $0.5558$ & $\pm 1.1537$ & $\pm 1.6115$ \\
        $\Real(A_{(a_1(1640)\to f_0(980) \pi)\KS})$ & \multicolumn{3}{c|}{---} & $-0.3569$ & ${}_{-0.7193}^{+0.6560}$ & ${}^{+1.1383}_{-1.1493}$ \\
        $\Imag(A_{(a_1(1640)\to f_0(980) \pi)\KS})$ & \multicolumn{3}{c|}{---} & $-1.7988$ & ${}_{-0.6458}^{+0.6287}$ & ${}^{+1.1608}_{-1.1620}$ \\
        \hline\hline
        \end{tabular}
        }
    \label{tab:fit-results}
\end{table}  
\endgroup   

\clearpage

\section*{Supplemental material}
\label{app:supplemental}

The twenty-two amplitudes included in the baseline fit are detailed in Table~\ref{tab:components}, showing the decay types, the notations of the couplings and angular distributions. The type $V$($S$) represents a state with spin $1$($0$).
For $\Bp\to a_1^+\KS$ decays, the topology is different from other two-body decays and the helicity angles are redefined, where
$\theta_{\rho\pi}$ ($\theta_{f_0\pi}$) is the angle between the direction of $\rhoz$ ($f_0$) in the $\rhoz\pip$ ($f_0\pip$) rest frame and the direction of $a_1^+$ in the $\Bp$ rest frame;
$\theta_{\pi\pi}$ is the angle between the direction of $\pip$ in the $\pip\pim$ rest frame and the direction of $\rhoz$ in the $a_1^+$ rest frame;
$\phi_{a_1}$ is the angle between the $\rhoz\pip$ and $\pip\pim$ decay planes.

\begin{table}[b]
    \centering
    \caption{Amplitudes contributing to the baseline amplitude  model. The first (second) letter in ``Type'' denotes the spin of the $\pip\pim$ ($\KS\pip$) system. }
    \begin{tabular}{cclc}
    \hline\hline
    $\mathrm{i}$ & Type & $A_i$ & $g_i\left(\theta_{\pi\pi}, \theta_{\KS\pi}, \phi\right)$  \\
    \hline 1 & & $A_{\rho K^*}^0$ & $\cos \theta_{\pi\pi} \cos \theta_{\KS\pi}$  \\
    2 & $V_1 V$ & $A_{\rho K^*}^{\parallel}$ & $\frac{1}{\sqrt{2}} \sin \theta_{\pi\pi} \sin \theta_{\KS\pi} \cos \phi$  \\
    3 & & $A_{\rho K^*}^{\perp}$ & $\frac{i}{\sqrt{2}} \sin \theta_{\pi\pi} \sin \theta_{\KS\pi} \sin \phi$  \\
    \hline 4 & & $A_{\omega K^*}^0$ & $\cos \theta_{\pi\pi} \cos \theta_{\KS\pi}$  \\
    5 & $V_2 V$ & $A_{\omega K^*}^{\parallel}$ & $\frac{1}{\sqrt{2}} \sin \theta_{\pi\pi} \sin \theta_{\KS\pi} \cos \phi$  \\
    6 & & $A_{\omega K^*}^{\perp}$ & $\frac{i}{\sqrt{2}} \sin \theta_{\pi\pi} \sin \theta_{\KS\pi} \sin \phi$  \\
    \hline 7 & $V_1 S$ & $A_{\rho(\KS\pi)_\text{S}}^0$ & $\frac{1}{\sqrt{3}} \cos \theta_{\pi\pi}$  \\
    8 & $V_2 S$ & $A_{\omega(\KS\pi)_\text{S}}^0$ & $\frac{1}{\sqrt{3}} \cos \theta_{\pi\pi}$ \\
    \hline 9 & $S_1 V$ & $A_{f_0(500) K^*}^0$ & $\frac{1}{\sqrt{3}} \cos \theta_{\KS\pi}$  \\
    10 & $S_2 V$ & $A_{f_0(980) K^*}^0$ & $\frac{1}{\sqrt{3}} \cos \theta_{\KS\pi}$  \\
    11 & $S_3 V$ & $A_{f_0(1370) K^*}^0$ & $\frac{1}{\sqrt{3}} \cos \theta_{\KS\pi}$  \\
    \hline 12 & $S_1 S$ & $A_{f_0(500)(\KS\pi)_\text{S}}^0$ & $\frac{1}{3}$  \\
    13 & $S_2 S$ & $A_{f_0(980)(\KS\pi)_\text{S}}^0$ & $\frac{1}{3}$  \\
    14 & $S_3 S$ & $A_{f_0(1370)(\KS\pi)_\text{S}}^0$ & $\frac{1}{3}$ \\
    \hline
    15 & $V_3 S$ & $A_{a_1(1260)(\to \rho \pi)\KS}^0$ & $\cos \theta_{\rho\pi} \cos \theta_{\pi\pi}$ \\
    16 & $V_3 S$ & $A_{a_1(1260)(\to \rho \pi)\KS}^\parallel$ & $\frac{1}{\sqrt{2}} \sin \theta_{\rho\pi} \sin \theta_{\pi\pi} \cos \phi_{a_1}$ \\
    17 & $V_3 S$ & $A_{a_1(1260)(\to f_0(500) \pi)\KS}$ & $\frac{1}{\sqrt{3}} \cos \theta_{f_0\pi}$ \\
    18 & $V_3 S$ & $A_{a_1(1260)(\to f_0(980) \pi)\KS}$ & $\frac{1}{\sqrt{3}} \cos \theta_{f_0\pi}$ \\
    \hline
    19 & $V_4 S$ & $A_{a_1(1640)(\to \rho \pi)\KS}^0$ & $\cos \theta_{\rho\pi} \cos \theta_{\pi\pi}$ \\
    20 & $V_4 S$ & $A_{a_1(1640)(\to \rho \pi)\KS}^\parallel$ & $\frac{1}{\sqrt{2}} \sin \theta_{\rho\pi} \sin \theta_{\pi\pi} \cos \phi_{a_1}$ \\
    21 & $V_4 S$ & $A_{a_1(1640)(\to f_0(500) \pi)\KS}$ & $\frac{1}{\sqrt{3}} \cos \theta_{f_0\pi}$ \\
    22 & $V_4 S$ & $A_{a_1(1640)(\to f_0(980) \pi)\KS}$ & $\frac{1}{\sqrt{3}} \cos \theta_{f_0\pi}$ \\
    \hline\hline
    \end{tabular}
    \label{tab:components}
\end{table}

The mass models selected for each resonance, along with the corresponding values of the constants used in these functions, are summarized in Table~\ref{tab:resonances}.
Most values are taken directly from Ref.~\cite{PDG2024}. However, for the $f_0(500), f_0(980), f_0(1370)$ and $a_1(1260)^+$ states, which exhibit a broad range of measured values, the values adopted in the baseline model are based on a combined assessment of the Ref.~\cite{PDG2024} and previous LHCb analyses~\cite{LHCb-PAPER-2018-042}. These parameters are varied within their uncertainties when evaluating the systematic uncertainties.

\begin{table}[tb]
    \centering
    \caption{Mass models and corresponding parameter values for each resonance. 
    The parameters $m_0$, $\Gamma_0$ and $d_\text{R}$ are the known mass, natural width and effective radius of each resonance, respectively.
    The values of these parameters are taken from Refs.~\cite{LHCb-PAPER-2018-042,PDG2024}.}
    \begin{tabular}{c c r c}
    \hline\hline
     Resonance & Lineshape & Parameter & Value \\
     \hline
     \multirow{3}{*}{$\rho(770)^0$} & \multirow{3}{*}{Gounaris--Sakurai~\cite{Gounaris:1968mw}} & $m_0$ $[\mevcc]$ &  $\phantom{-0}769.0\phantom{00}$  \\
      & & $\Gamma_0$ $[\mevcc]$ & $\phantom{-0}150.9\phantom{00}$  \\
      & & $d_\text{R}$ $[\gev^{-1}]$ & $\phantom{-000}5.3\phantom{00}$ \\
    \hline
     \multirow{3}{*}{$\omega(782)$} & \multirow{3}{*}{Relativistic Breit--Wigner} & $m_0$ $[\mevcc]$ &  $\phantom{-0}782.66\phantom{0}$ \\
      & & $\Gamma_0$ $[\mevcc]$ & $\phantom{-000}8.68\phantom{0}$\\
      & & $d_\text{R}$ $[\gev^{-1}]$ & $\phantom{-000}3.0\phantom{00}$ \\
    \hline
     \multirow{2}{*}{$f_0(500)$} & \multirow{2}{*}{Relativistic Breit--Wigner} & $m_0$ $[\mevcc]$ &  $\phantom{-0}513\phantom{.000}$  \\
      & & $\Gamma_0$ $[\mevcc]$ & $\phantom{-0}335\phantom{.000}$  \\
    \hline
     \multirow{4}{*}{$f_0(980)$} & \multirow{4}{*}{Flatté~\cite{Flatte:1976xu,Flatte:1976xv,Bugg:2008ig}} & $m_0$ $[\mevcc]$ &  $\phantom{-0}965\phantom{.000}$  \\
      & & $\alpha$ $[\gev^{-2}]$ & $\phantom{-000}2.0\phantom{00}$ \\
      & & $g_{\pi\pi}$ $[\gev]$ & float (to be $0.16\pm0.01$)\\
      & & $g_{KK}/g_{\pi\pi}$ & float (to be $3.06\pm0.08$)  \\
    \hline
     \multirow{2}{*}{$f_0(1370)$} & \multirow{2}{*}{Relativistic Breit--Wigner} & $m_0$ $[\mevcc]$ &  $\phantom{-}1350\phantom{.000}$ \\
      & & $\Gamma_0$ $[\mevcc]$ & $\phantom{-0}350\phantom{.000}$  \\
    \hline
     \multirow{3}{*}{$\Kstar(892)^+$} & \multirow{3}{*}{Relativistic Breit--Wigner} & $m_0$ $[\mevcc]$ & $\phantom{-0}891.67\phantom{0}$  \\
      & & $\Gamma_0$ $[\mevcc]$ & $\phantom{-00}51.4\phantom{00}$  \\
      & & $d_\text{R}$ $[\gev^{-1}]$ & $\phantom{-000}3.0\phantom{00}$ \\
    \hline
     \multirow{5}{*}{$(\KS\pip)_\text{S}$} & \multirow{5}{*}{LASS~\cite{Aston:1987ir}} & $m_{K^*_0(1430)}$ $[\mevcc]$ & $\phantom{-}1425\phantom{.000}$  \\
      & & $\Gamma_{K^*_0(1430)}$ $[\mevcc]$ & $\phantom{-0}270\phantom{.000}$  \\
      & & FF1 & $\phantom{000}$$-0.707$$\phantom{0}$ \\
      & & $a$ $[\gev^{-1}]$ & float (to be $3.15\pm0.20$)\\
      & & $b$ $[\gev^{-1}]$ & float  (to be $1.00\pm0.12$)\\
    \hline
      \multirow{3}{*}{$a_1(1260)^+$} & \multirow{3}{*}{Relativistic Breit--Wigner} & $m_0$ $[\mevcc]$ &  $\phantom{-}1230\phantom{.000}$ \\
       & & $\Gamma_0$ $[\mevcc]$ & $\phantom{-0}420\phantom{.000}$  \\
       & & $d_\text{R}$ $[\gev^{-1}]$ & $\phantom{-000}3.0\phantom{00}$ \\
      \hline
      \multirow{3}{*}{$a_1(1640)^+$} & \multirow{3}{*}{Relativistic Breit--Wigner} & $m_0$ $[\mevcc]$ &  $\phantom{-}1655\phantom{.000}$ \\
       & & $\Gamma_0$ $[\mevcc]$ & $\phantom{-0}250\phantom{.000}$  \\
       & & $d_\text{R}$ $[\gev^{-1}]$ & $\phantom{-000}3.0\phantom{00}$ \\
    \hline\hline
    \end{tabular}
    \label{tab:resonances}
\end{table}

\clearpage
Several two-dimensional plots of the mass and angular variables are shown below.
The plots use candidates from the signal region, $5.24<m_{\pip\pim\KS\pip}<5.32\gevcc$, which contains about 6\% background contamination.

\begin{figure}[b]
    \centering
    \includegraphics[width=0.43\linewidth]{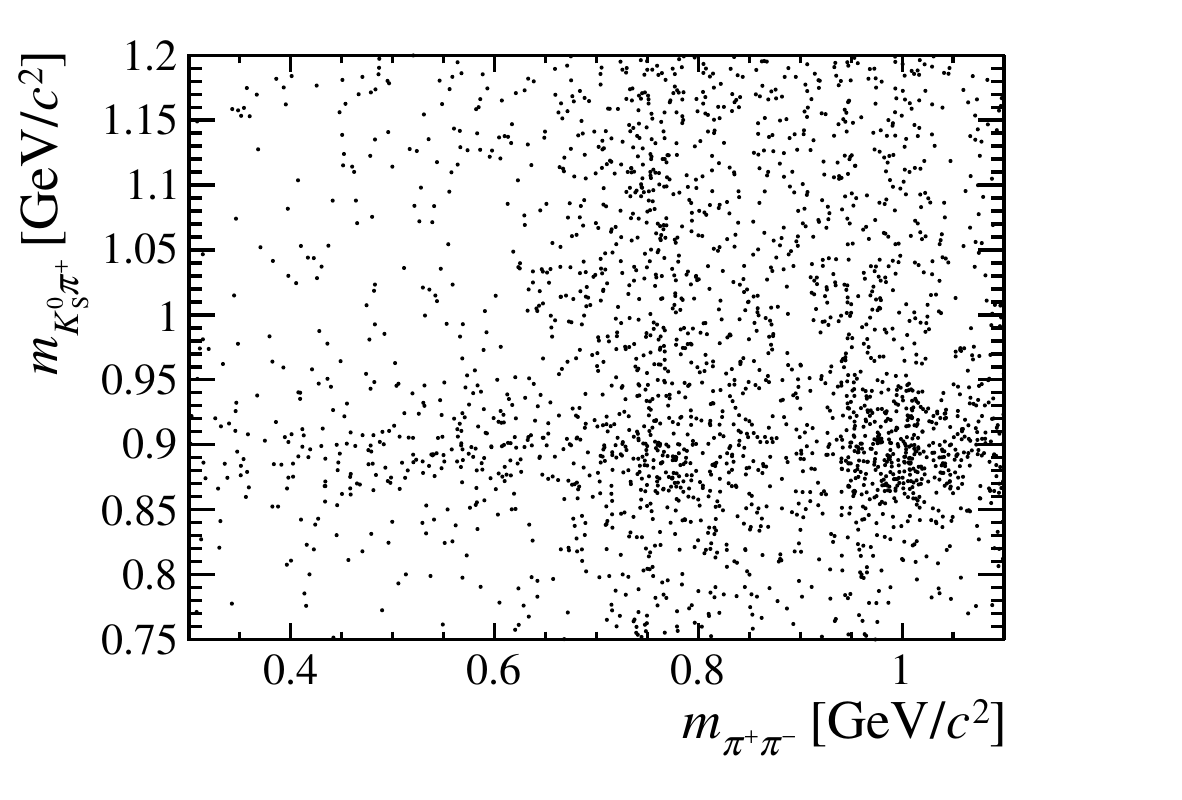}
    \includegraphics[width=0.43\linewidth]{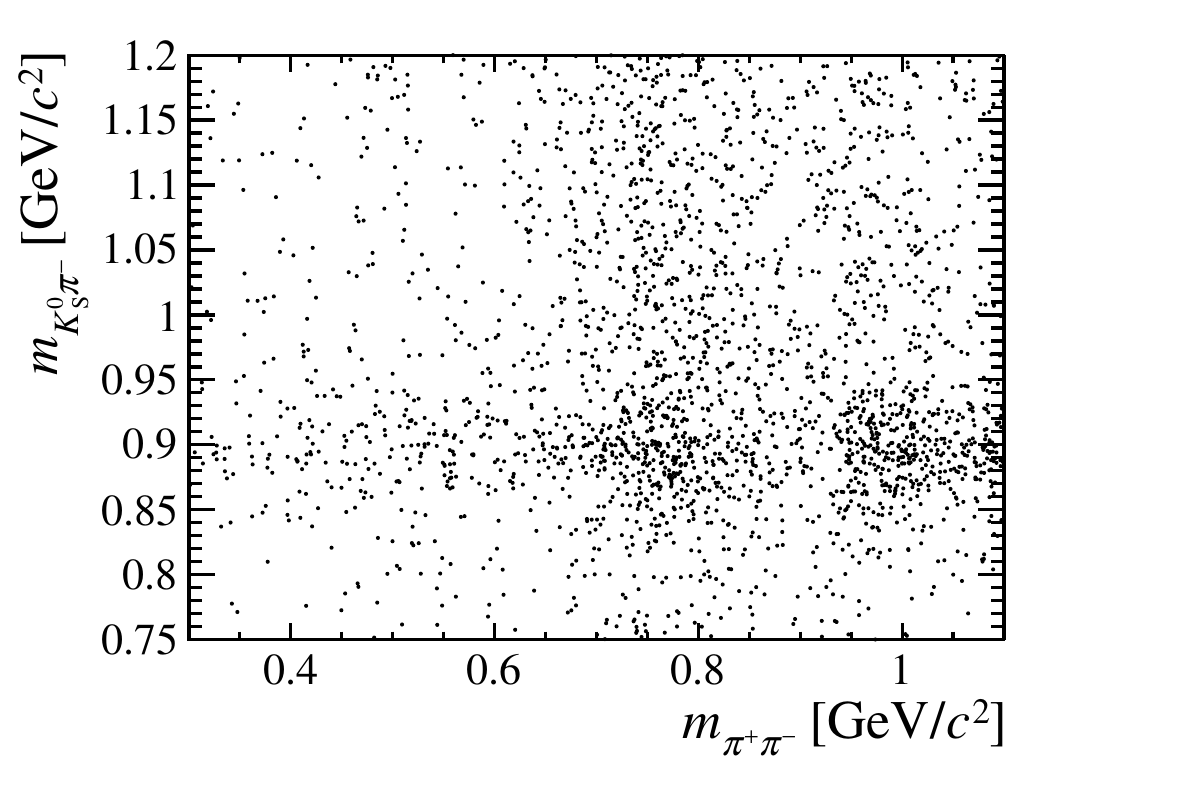}
    \caption{Scatter plots of $m_{\pip\pim}$ \vs $m_{\KS\pipm}$ for (left) $\Bp$ and (right) $\Bm$ decays. The plots use candidates from the signal region, which contains around 6\% background contamination.}
    \label{fig:mRho-mKst}
\end{figure}

Figure~\ref{fig:mRho-mKst} shows the distributions of $m_{\pip\pim}$ \vs $m_{\KS\pip}$, where clear vertical bands corresponding to the $\rho(770)^0$ and $f_0(980)$ resonances, as well as a horizontal band associated with the $K^*(892)^+$, can be seen.

\begin{figure}[b]
    \centering
    \includegraphics[width=0.43\linewidth]{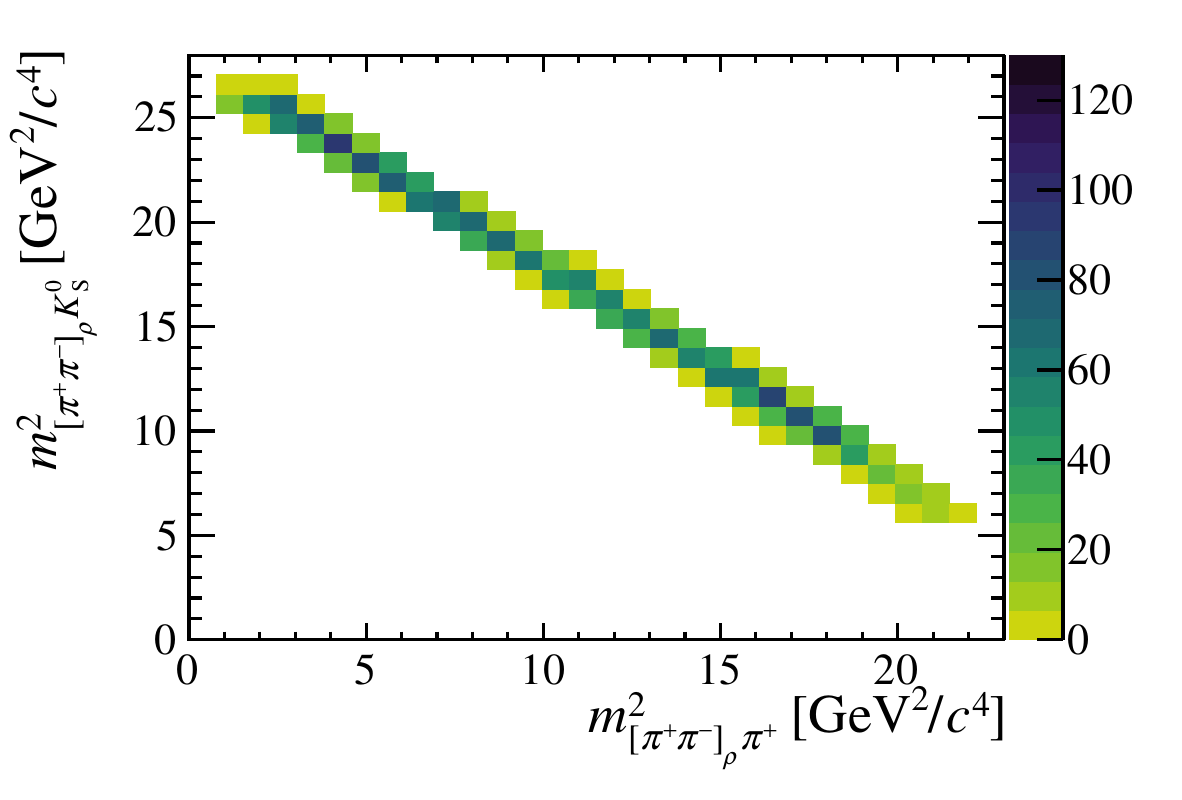}
    \includegraphics[width=0.43\linewidth]{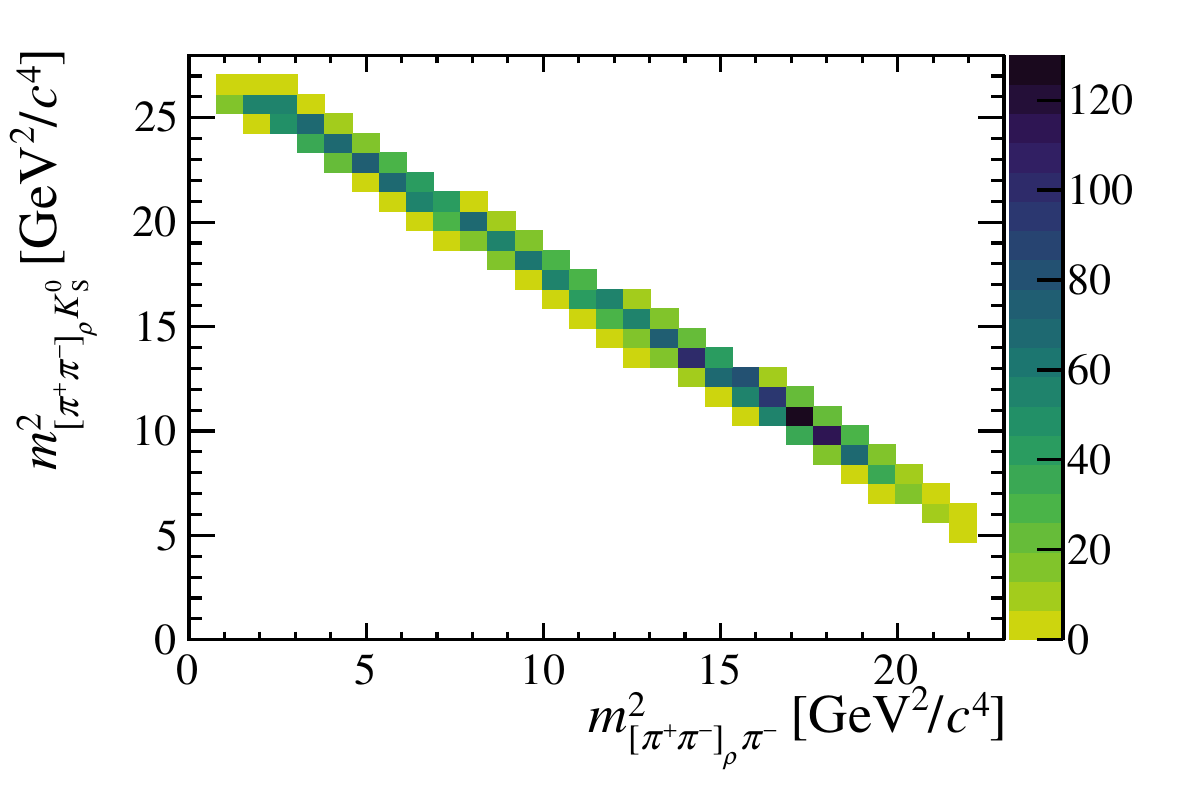}
    \includegraphics[width=0.43\linewidth]{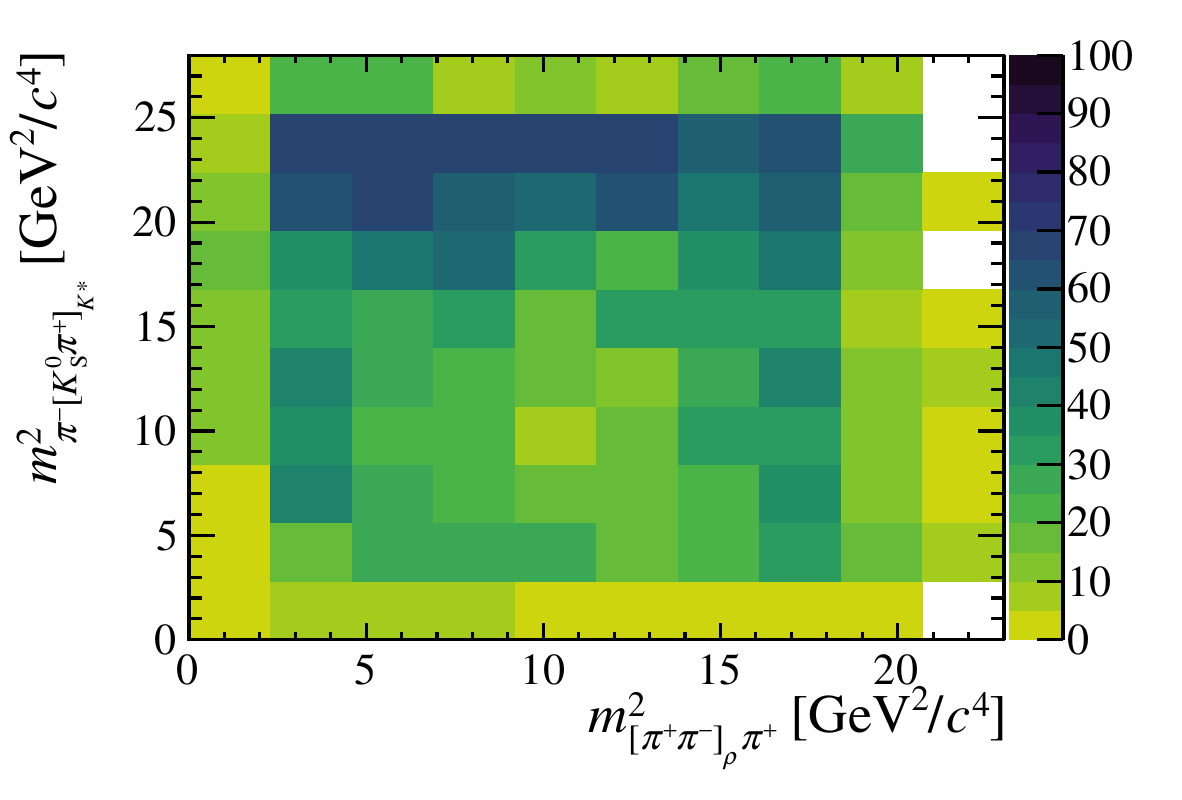}
    \includegraphics[width=0.43\linewidth]{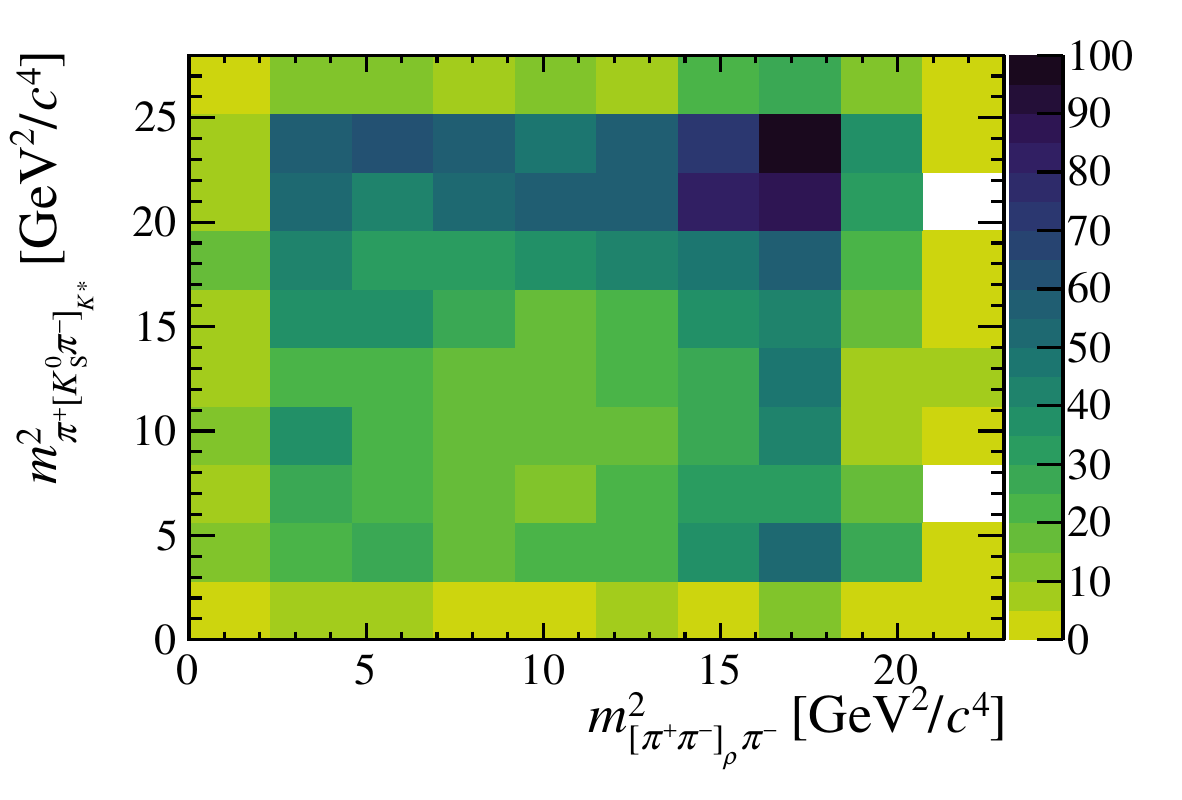}
    \caption{Plots of (top) $m^2_{[\pip\pim]_{\rho}\pipm}$ \vs $m^2_{[\pip\pim]_{\rho}\KS}$, and (bottom) $m^2_{[\pip\pim]_{\rho}\pipm}$ \vs $m^2_{\pimp[\KS\pipm]_{K^*}}$. The subscripts indicate particles originate from the $\rho^0$ or $\Kstar$ resonances. The plots correspond to (left) $\Bp$ and (right) $\Bm$ decays, using candidates from the signal region with around 6\% background contamination.}
    \label{fig:m3pi-mKpipi}
\end{figure}

Figure~\ref{fig:m3pi-mKpipi} shows, in the top the plots of $m^2_{[\pip\pim]_{\rho}\pip}$ \vs $m^2_{[\pip\pim]_{\rho}\KS}$, and in the bottom $m^2_{[\pip\pim]_{\rho}\pip}$ \vs $m^2_{\pim[\KS\pip]_{K^*}}$.
The subscripts indicate that the particles originate from the $\rho^0$ or $\Kstarp$ resonances.
Note that they are not Dalitz plots due to selections applied on $m_{\pip\pim}$ and $m_{\KS\pip}$ in this analysis.
No clear $a_1\to\pi\pi\pi$ or $K_1\to K \pi\pi$ structures are visible, as these modes have relatively small fractions. However, the amplitude fit confirms that the $a_1$ contributions are necessary, while the $K_1$ contributions are negligible.

\begin{figure}[b]
    \centering
    \includegraphics[width=0.43\linewidth]{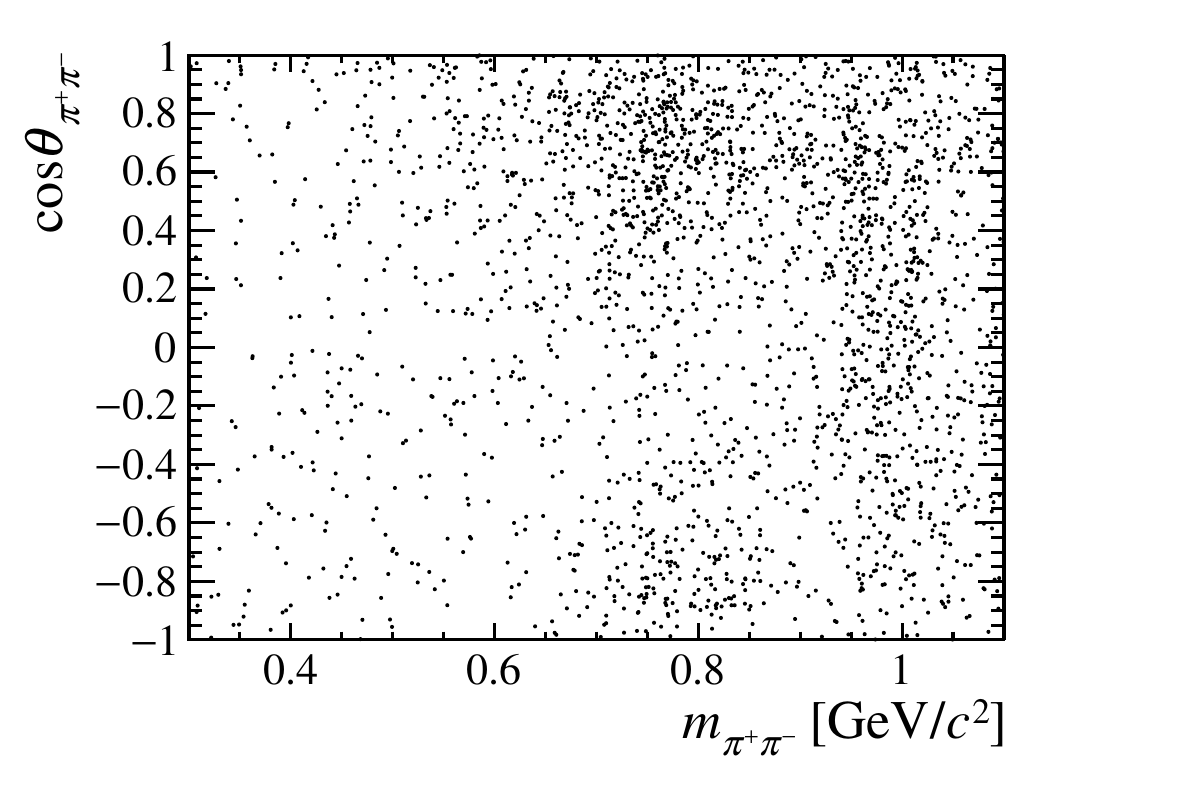}
    \includegraphics[width=0.43\linewidth]{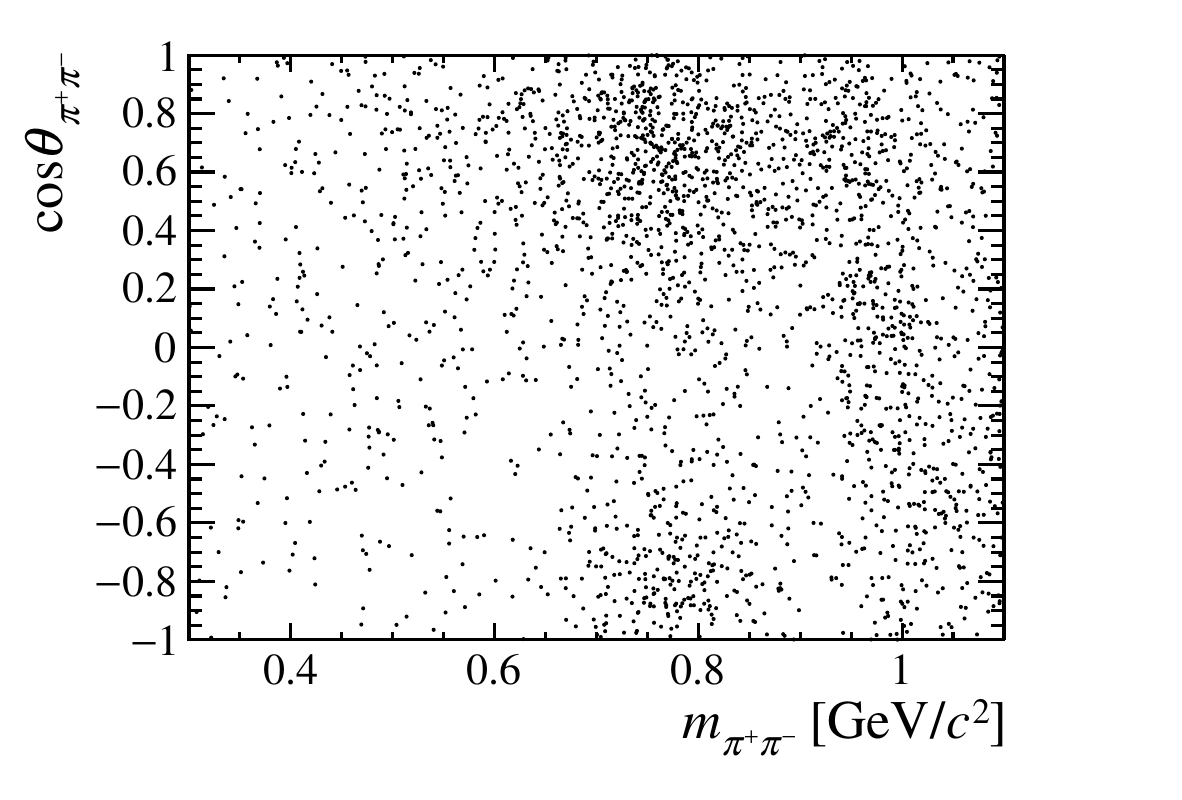}
    \includegraphics[width=0.43\linewidth]{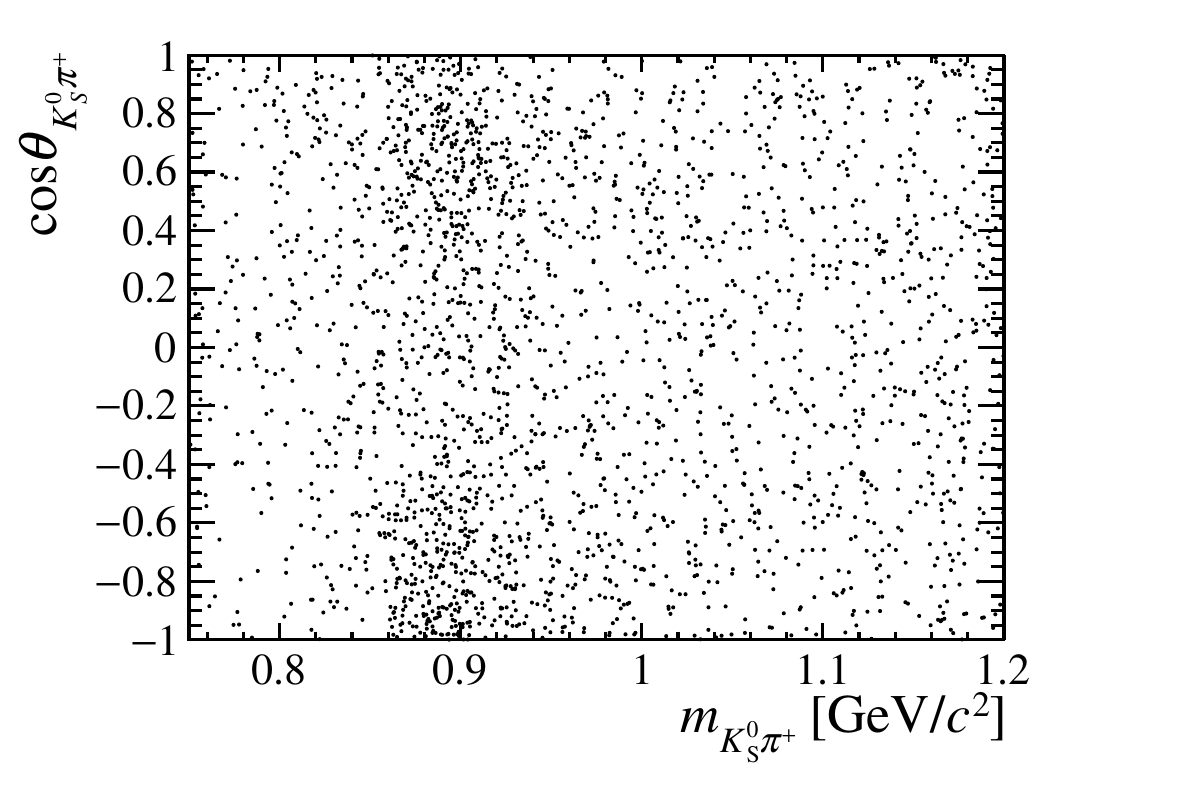}
    \includegraphics[width=0.43\linewidth]{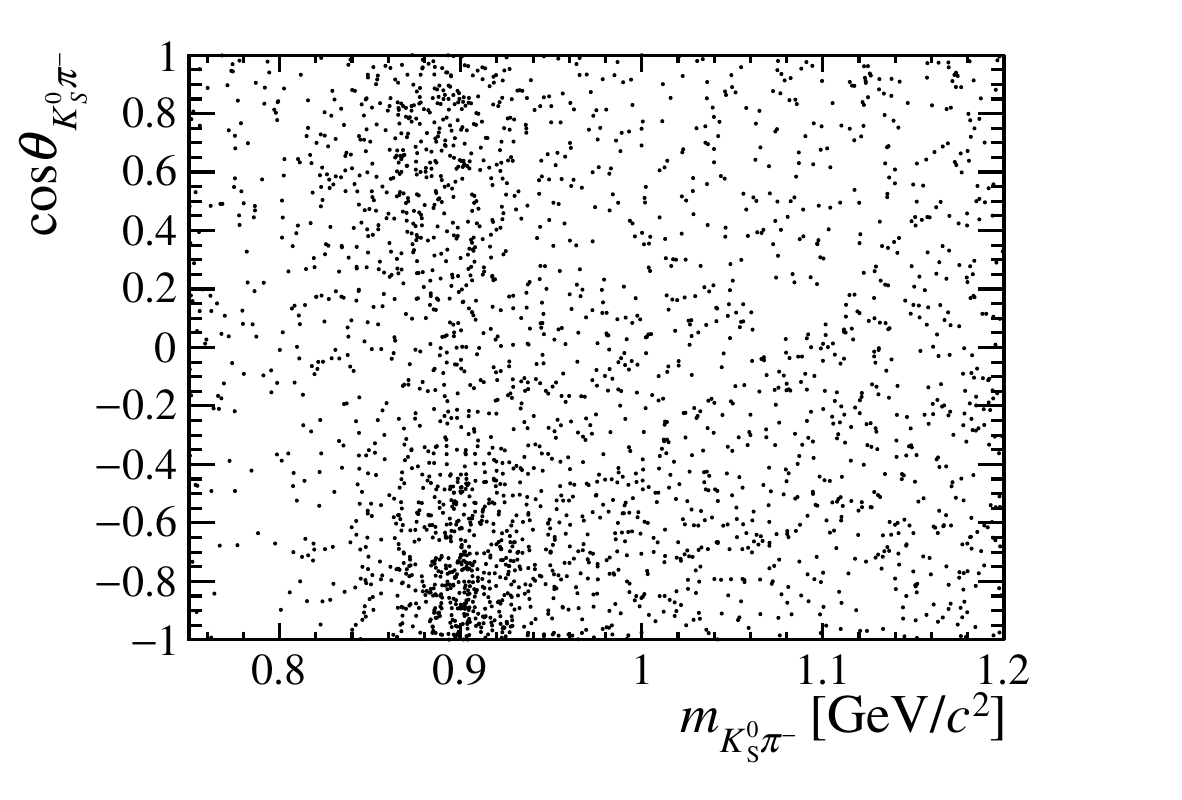}
    \caption{Scatter plots of (top) $m_{\pip\pim}$s \vs $\cos\theta_{\pip\pim}$, and (bottom) $m_{\KS\pipm}$ \vs $\cos\theta_{\KS\pipm}$. The plots correspond to (left) $\Bp$ and (right) $\Bm$ decays. Candidates are taken from the signal region, which contains around 6\% background contamination.}
    \label{fig:m-costheta}
\end{figure}

Figure~\ref{fig:m-costheta} presents the correlations between mass and angular variables, which are $m_{\pip\pim}$ \vs $\cos\theta_{\pip\pim}$ and $m_{\KS\pip}$ \vs $\cos\theta_{\KS\pip}$. 
In the top plots, clear P-wave structures are seen around $m_{\pip\pim}=0.8\gevcc$ and S-wave structures around $1.0\gevcc$. 
In the bottom plots, the P-wave contribution is visible near $m_{\KS\pip}=0.9\gevcc$.

\clearpage
\bibliographystyle{LHCb}
\bibliography{main,standard,LHCb-PAPER,LHCb-CONF,LHCb-DP,LHCb-TDR}
 
\newpage
\centerline
{\large\bf LHCb collaboration}
\begin
{flushleft}
\small
R.~Aaij$^{38}$\lhcborcid{0000-0003-0533-1952},
A.S.W.~Abdelmotteleb$^{57}$\lhcborcid{0000-0001-7905-0542},
C.~Abellan~Beteta$^{51}$\lhcborcid{0009-0009-0869-6798},
F.~Abudin{\'e}n$^{57}$\lhcborcid{0000-0002-6737-3528},
T.~Ackernley$^{61}$\lhcborcid{0000-0002-5951-3498},
A. A. ~Adefisoye$^{69}$\lhcborcid{0000-0003-2448-1550},
B.~Adeva$^{47}$\lhcborcid{0000-0001-9756-3712},
M.~Adinolfi$^{55}$\lhcborcid{0000-0002-1326-1264},
P.~Adlarson$^{85}$\lhcborcid{0000-0001-6280-3851},
C.~Agapopoulou$^{14}$\lhcborcid{0000-0002-2368-0147}
% ...

{\footnotesize \it

% ...
$^{14}$Universit{\'e} Paris-Saclay, CNRS/IN2P3, IJCLab, Orsay, France\\
% ...
$^{38}$Nikhef National Institute for Subatomic Physics, Amsterdam, Netherlands\\
% ...
$^{47}$Instituto Galego de F{\'\i}sica de Altas Enerx{\'\i}as (IGFAE), Universidade de Santiago de Compostela, Santiago de Compostela, Spain\\
% ...
$^{51}$Physik-Institut, Universit{\"a}t Z{\"u}rich, Z{\"u}rich, Switzerland\\
% ...
$^{55}$H.H. Wills Physics Laboratory, University of Bristol, Bristol, United Kingdom\\
% ...
$^{57}$Department of Physics, University of Warwick, Coventry, United Kingdom\\
% ...
$^{60}$School of Physics and Astronomy, University of Glasgow, Glasgow, United Kingdom\\
$^{61}$Oliver Lodge Laboratory, University of Liverpool, Liverpool, United Kingdom\\
% ...
$^{69}$Syracuse University, Syracuse, NY, United States\\
% ...
$^{85}$Department of Physics and Astronomy, Uppsala University, Uppsala, Sweden, associated to $^{60}$\\
% ...
}
\end{flushleft}

\end{document}